\documentclass{article}

\usepackage{arxiv}

\usepackage{natbib}
\usepackage[utf8]{inputenc} 
\usepackage[T1]{fontenc}    
\usepackage[hidelinks]{hyperref}
\usepackage{url}            
\usepackage{booktabs}       
\usepackage{amsfonts}       
\usepackage{nicefrac}       
\usepackage{microtype}      
\usepackage{lipsum}
\usepackage{graphicx}
\graphicspath{ {./images/} }

\usepackage{soul}
\sethlcolor{yellow}

\usepackage[section]{algorithm}
\usepackage[noend]{algorithmic}
\usepackage{adjustbox}

\usepackage{bm}
\usepackage[font = small,skip = 4pt]{caption} 
\usepackage[shortlabels]{enumitem}
\usepackage[acronym]{glossaries}
\usepackage{physics}
\usepackage{tabularx}
\usepackage{stackengine}
\usepackage{tikz}

\usepackage{multirow}
\usepackage{pgfplots}

\usepackage[capitalize, noabbrev]{cleveref}
\usepackage{subcaption} 
\usepackage{xspace}

\usepackage{multirow}
\usepackage{tabularx}
\usepackage[most]{tcolorbox}
\usepackage{wrapfig}
\usepackage{booktabs}

\usepackage{placeins}
\usepackage{needspace}

\let\originalleft\left
\let\originalright\right
\renewcommand{\left}{\mathopen{}\mathclose\bgroup\originalleft}
\renewcommand{\right}{\aftergroup\egroup\originalright}

\newcommand\eqfor{\quad\text{for }}
\newcommand\eqforall{\quad\text{for all }}

\makeglossaries

\newacronym{sbgm}{SBGM}{score-based generative modeling}

\newacronym{osm}{OSM}{ordinary score matching}
\newacronym{dsm}{DSM}{Denoising Score Matching}
\newacronym{smld}{SMLD}{Score Matching with Langevin Dynamics}
\newacronym{ddpm}{DDPM}{Denoising Diffusion Probabilistic Models}
\newacronym{sdesbm}{SDE-driven SBGM}{SDE-driven score-based diffusion model}

\newacronym{ode}{ODE}{ordinary differential equation}
\newacronym{pde}{PDE}{partial differential equation}
\newacronym{sde}{SDE}{stochastic differential equation}
\newacronym{spde}{SPDE}{stochastic partial differential equation}

\newacronym{sgd}{SGD}{stochastic gradient descent}

\newacronym{ula}{ULA}{unadjusted Langevin algorithm}

\newacronym{ddbm}{DDBM}{Denoising Diffusion Bridge Models}
\newacronym{fm}{FM}{Flow Matching}

\newacronym{fid}{FID}{Fréchet inception distance}
\newacronym{is}{IS}{inception score}
\newacronym{kid}{KID}{kernel inception distance}

\glsdisablehyper

\newcommand\declaresymbol[2]{\newcommand{#1}{\TextOrMath{$#2$\xspace}{#2}}}

\declaresymbol\probabilityspace\Omega
\declaresymbol\eventsystem{\mathcal A}
\declaresymbol\probabilitymeasure{\operatorname P}
\declaresymbol\expectation{\operatorname E}

\declaresymbol\uniformdistribution{\mathcal U}

\declaresymbol\lebesguemeasure\lambda
\declaresymbol\forwarddistribution\pi
\declaresymbol\forwarddensity p
\declaresymbol\datadistribution\mu
\declaresymbol\targetdensity\forwarddensity
\declaresymbol\lossmeasure\eta

\declaresymbol\loss L
\declaresymbol\weight\zeta

\declaresymbol\langevinprocess X
\declaresymbol\langevindrift\eta
\declaresymbol\langevinantisymmetricdrift K
\declaresymbol\langevindiffusioncoefficient\epsilon
\declaresymbol\langevincovariance{\mathrm E}

\declaresymbol\discretizedlangevinprocess Y

\declaresymbol\drift b
\declaresymbol\diffusioncoefficient\sigma
\declaresymbol\wienerprocess W
\declaresymbol\evolutionprocess U
\declaresymbol\forwardprocess\evolutionprocess

\declaresymbol\wienercovariance Q
\declaresymbol\wienercovariancedensity{\MakeLowercase\wienercovariance}
\declaresymbol\wienercovariancecorrelationlength\ell
\declaresymbol\evolutioncovariance\Sigma 

\declaresymbol\timedomain I

\declaresymbol\score s

\declaresymbol\integrand f

\declaresymbol\timepoint t
\declaresymbol\prevtimepoint s
\declaresymbol\evolutionpoint{\MakeLowercase\evolutionprocess}
\declaresymbol\noisemode v
\declaresymbol\sample\evolutionpoint

\declaresymbol\timepointmax T
\declaresymbol\timestepcontinuous {dt}
\declaresymbol\reversetimestepcontinuous {dt_{rev}}

\declaresymbol\dimension d
\declaresymbol\domain\Lambda
\declaresymbol\point x
\declaresymbol\otherpoint y
\declaresymbol\measurableset B
\declaresymbol\measurabletimeset A

\declaresymbol\tamingcoefficient\gamma
\declaresymbol\sobolevspaceorder r
\declaresymbol\powerorder r
\declaresymbol\exponentialparam k

\declaresymbol\anisotropycoefficient g

\declaresymbol\discretedimension d
\declaresymbol\discretedomain D
\declaresymbol\indexsetelement i

\declaresymbol\sequenceindex i
\declaresymbol\sequenceindexmax k

\declaresymbol\transition\tau
\declaresymbol\diffusivitytransition\alpha
\declaresymbol\anisotropytransition\lambda

\declaresymbol\timestep k
\declaresymbol\gaussianvariable\xi

\newcommand\subscriptindex[2]{#1,\;#2}
\newcommand\subscriptmin[2]{#1\;\wedge\;#2}
\newcommand\subscriptmax[2]{#1\;\vee\;#2}

\newcommand\dcc[2]{#1_#2}
\newcommand\dbb[2]{#1_{\subscriptindex{#2_1-1}{#2_2-1}}}
\newcommand\dbc[2]{#1_{\subscriptindex{#2_1-1}{#2_2}}}
\newcommand\dbf[2]{#1_{\subscriptindex{#2_1-1}{#2_2+1}}}
\newcommand\dfc[2]{#1_{\subscriptindex{#2_1+1}{#2_2}}}
\newcommand\dcb[2]{#1_{\subscriptindex{#2_1}{#2_2-1}}}
\newcommand\dcf[2]{#1_{\subscriptindex{#2_1}{#2_2+1}}}
\newcommand\dfb[2]{#1_{\subscriptindex{#2_1+1}{#2_2-1}}}
\newcommand\dff[2]{#1_{\subscriptindex{#2_1+1}{#2_2+1}}}

\makeatletter
\newcommand{\setword}[2]{%
    \phantomsection
    #1\def\@currentlabel{\unexpanded{#1}}\label{#2}%
}
\makeatother

\crefname{paragraph}{Paragraph}{Paragraphs}

\creflabelformat{equation}{#2(#1)#3}
\crefname{equation}{}{}
\Crefname{equation}{}{}

\title{Score-Based Generative Modeling through\\Anisotropic Stochastic Partial Differential Equations}

\author{
  Sascha Holl\thanks{Corresponding author.}\\
  Max Planck Institute for Informatics\\
  Saarland University\\
  Saarbrücken, Germany\\
  \texttt{sascha.holl@gmail.com}\\
  \newline
  \And
  Jente Vandersanden\\
  Max Planck Institute for Informatics\\
  Saarland University\\
  Saarbrücken, Germany\\
  \texttt{jvanders@mpi-inf.mpg.de}\\
  \And
  Gurprit Singh\\
  Advanced Micro Devices, Inc. (AMD)\\
  Munich, Germany\\
  \texttt{gurprit.singh@amd.com}\\
  \And
  Hans-Peter Seidel\\
  Max Planck Institute for Informatics\\
  Saarbrücken, Germany\\
  \texttt{hpseidel@mpi-inf.mpg.de}\\
}

\begin{document}

\maketitle

\begin{figure*}[t]
    \centering
    \newcommand{\vcenterlabel}[1]{%
  \rotatebox{90}{\small #1}%
}

\noindent
\begin{tabular}{@{\hspace{-1.2cm}}p{.55\textwidth}@{\hspace{-10pt}}p{.4\textwidth}@{}}
    \small\newcommand{\PlotSingleImage}[1]{%
    \begin{scope}
        \clip (0,0) -- (2.5,0) -- (2.5,2.5) -- (0,2.5) -- cycle;
        \path[fill overzoom image=figures/fwd_process_vis/#1] (0,0) rectangle (2.5cm,2.5cm);
    \end{scope}
    \draw (0,0) -- (2.5,0) -- (2.5,2.5) -- (0,2.5) -- cycle;
}

\newcommand\scalevalue{0.9}    
\newcommand\scalevalueBigger{.71}    

\small
\def\arraystretch{.4}
\setlength{\tabcolsep}{.03cm}
\begin{tabular}{ccccccccccc}
~ &
$t=0$ &
$t=\frac14T$ &
$t=\frac24T$ &
$t=\frac34T$ &
$t=T$
\vspace{.5mm}
\\
\rotatebox{90}{\hspace{2mm}\tiny\def\stackalignment{l}\stackanchor{No drift +}{isotropic noise}}
&
\begin{tikzpicture}[scale=\scalevalueBigger]
\PlotSingleImage{iso_noise/0.jpg}
\end{tikzpicture}
&
\begin{tikzpicture}[scale=\scalevalueBigger]
\PlotSingleImage{iso_noise/2.jpg}
\end{tikzpicture}
&
\begin{tikzpicture}[scale=\scalevalueBigger]
\PlotSingleImage{iso_noise/4.jpg}
\end{tikzpicture}
&
\begin{tikzpicture}[scale=\scalevalueBigger]
\PlotSingleImage{iso_noise/6.jpg}
\end{tikzpicture}
&
\begin{tikzpicture}[scale=\scalevalueBigger]
\PlotSingleImage{iso_noise/8.jpg}
\end{tikzpicture}
\\
\rotatebox{90}{\hspace{1.5mm}\tiny\def\stackalignment{l}\stackanchor{No drift +}{anisotropic noise}}
&
\begin{tikzpicture}[scale=\scalevalueBigger]
\PlotSingleImage{aniso_noise/0.jpg}
\end{tikzpicture}
&
\begin{tikzpicture}[scale=\scalevalueBigger]
\PlotSingleImage{aniso_noise/2.jpg}
\end{tikzpicture}
&
\begin{tikzpicture}[scale=\scalevalueBigger]
\PlotSingleImage{aniso_noise/4.jpg}
\end{tikzpicture}
&
\begin{tikzpicture}[scale=\scalevalueBigger]
\PlotSingleImage{aniso_noise/6.jpg}
\end{tikzpicture}
&
\begin{tikzpicture}[scale=\scalevalueBigger]
\PlotSingleImage{aniso_noise/8.jpg}
\end{tikzpicture}
\\
\rotatebox{90}{\hspace{1.5mm}\tiny\def\stackalignment{l}\stackanchor{Isotropic drift +}{no noise}}
&
\begin{tikzpicture}[scale=\scalevalueBigger]
\PlotSingleImage{heat/0.jpg}
\end{tikzpicture}
&
\begin{tikzpicture}[scale=\scalevalueBigger]
\PlotSingleImage{heat/2.jpg}
\end{tikzpicture}
&
\begin{tikzpicture}[scale=\scalevalueBigger]
\PlotSingleImage{heat/4.jpg}
\end{tikzpicture}
&
\begin{tikzpicture}[scale=\scalevalueBigger]
\PlotSingleImage{heat/6.jpg}
\end{tikzpicture}
&
\begin{tikzpicture}[scale=\scalevalueBigger]
\PlotSingleImage{heat/8.jpg}
\end{tikzpicture}
\\
\rotatebox{90}{\hspace{1.25mm}\tiny\def\stackalignment{l}\stackanchor{Anisotropic drift +}{no noise}}
&
\begin{tikzpicture}[scale=\scalevalueBigger]
\PlotSingleImage{aniso_heat/0.jpg}
\end{tikzpicture}
&
\begin{tikzpicture}[scale=\scalevalueBigger]
\PlotSingleImage{aniso_heat/2.jpg}
\end{tikzpicture}
&
\begin{tikzpicture}[scale=\scalevalueBigger]
\PlotSingleImage{aniso_heat/4.jpg}
\end{tikzpicture}
&
\begin{tikzpicture}[scale=\scalevalueBigger]
\PlotSingleImage{aniso_heat/6.jpg}
\end{tikzpicture}
&
\begin{tikzpicture}[scale=\scalevalueBigger]
\PlotSingleImage{aniso_heat/8.jpg}
\end{tikzpicture}
\\
\rotatebox{90}{\hspace{1.25mm}\tiny\def\stackalignment{l}\stackanchor{Anisotropic drift +}{anisotropic noise}}
&
\begin{tikzpicture}[scale=\scalevalueBigger]
\PlotSingleImage{combination/0.jpg}
\end{tikzpicture}
&
\begin{tikzpicture}[scale=\scalevalueBigger]
\PlotSingleImage{combination/2.jpg}
\end{tikzpicture}
&
\begin{tikzpicture}[scale=\scalevalueBigger]
\PlotSingleImage{combination/4.jpg}
\end{tikzpicture}
&
\begin{tikzpicture}[scale=\scalevalueBigger]
\PlotSingleImage{combination/6.jpg}
\end{tikzpicture}
&
\begin{tikzpicture}[scale=\scalevalueBigger]
\PlotSingleImage{combination/8.jpg}
\end{tikzpicture}
\\
\end{tabular}

    &
    \resizebox{1.22\linewidth}{!}{%
        
\newcommand{\imgcell}[2][]{%
  \adjustbox{valign=m}{\includegraphics[#1]{#2}}%
}
\newcommand{\rowcap}[1]{%
  \rotatebox[origin=c]{90}{\footnotesize #1}%
}
\newcommand{\colcap}[1]{\footnotesize #1}

\setlength{\tabcolsep}{.05cm} 
\renewcommand{\arraystretch}{1}

\adjustbox{max width=\textwidth}{%
\begin{tabular}{ccc}

\imgcell[width=0.16\textwidth]{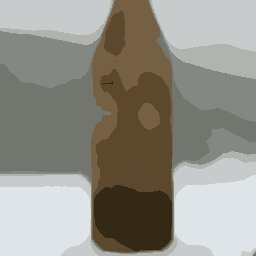}
& \imgcell[width=0.16\textwidth]{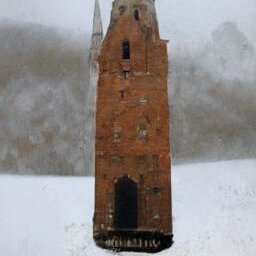}
& \imgcell[width=0.16\textwidth]{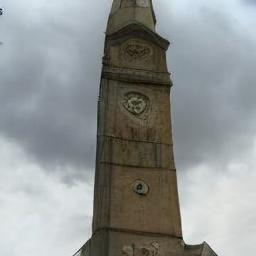}
\\\noalign{\vskip 3pt}

\imgcell[width=0.16\textwidth]{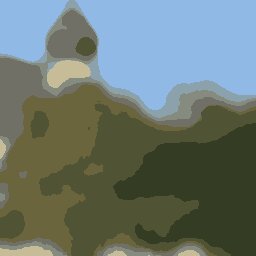}
& \imgcell[width=0.16\textwidth]{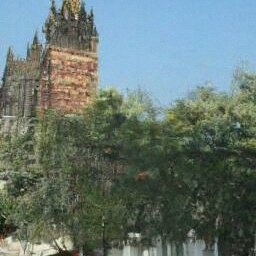}
& \imgcell[width=0.16\textwidth]{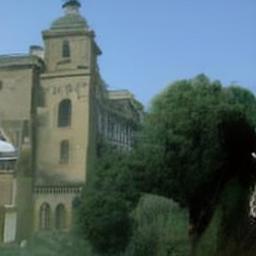}
\\\noalign{\vskip 3pt}

\imgcell[width=0.16\textwidth]{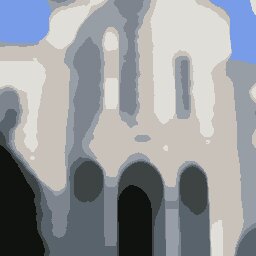}
& \imgcell[width=0.16\textwidth]{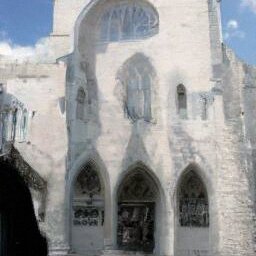}
& \imgcell[width=0.16\textwidth]{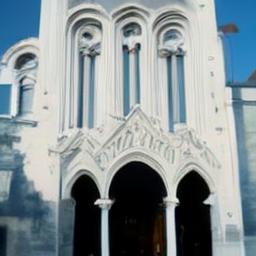}

\\\noalign{\vskip .1cm}

\footnotesize {\def\stackalignment{l}\stackanchor{Stroke painting}{}}
&
\footnotesize {\def\stackalignment{l}\stackanchor{SDEdit +}{isotropic baseline}}
&
\footnotesize {\def\stackalignment{l}\stackanchor{SDEdit +}{\emph{Ours (anisotropic)}}}
\\
\end{tabular}%
}

    }
\end{tabular}

    \caption{
        Our anisotropic diffusion framework preserves geometric features over longer time scales, thereby affecting reconstructability during generative sampling.
        Left: Visualization of our forward processes, decomposed into drift (deterministic destruction) and noise (stochastic destruction) components, at selected time points from $\timepoint=0$ up to the terminal time $\timepoint=\timepointmax$. From top to bottom:
        (1) isotropic noise without drift (as in traditional diffusion models),
        (2) anisotropic (edge-preserving) noise without drift,
        (3) isotropic drift (heat-equation–like blur) without noise,
        (4) anisotropic drift (edge-preserving blur) without noise, and
        (5) anisotropic drift combined with anisotropic noise.
        As $\timepoint\to\timepointmax$, a transition to isotropy is required since the backward generative process must start from fully destructed data, as discussed in \Cref{sec:residual-dependence}. This transition is intentionally omitted here to isolate the effect of the individual components. 
        Right: We demonstrate the geometry-awareness of the proposed approach on the stroke-to-image generation task \textsc{SDEdit}~\citep{meng2021sdedit}.
        The isotropic, \acrshort{sde}-based model of \citet{song2021scorebased} preserves the geometric structure of the input stroke image substantially less faithfully than our anisotropic, \acrshort{spde}-based model.
    }
    \label{fig:teaser}
\end{figure*}

\begin{abstract}
    \Acrlong{sbgm} has achieved state-of-the-art performance in image generation, with the quality of generated images being highly dependent on the design of the forward (diffusion) process. Among these, models based on \glspl{sde} have proven particularly effective. While traditional methods aim to progressively destroy all image information to enable reconstruction from pure noise, we propose a class of anisotropic \glspl{spde} that preserve the geometric structure of the data over longer time scales throughout the transformation. These \glspl{spde} consist of a drift term that enforces deterministic destruction via structured smoothing, and a diffusion coefficient that enables random destruction through noise injection. Both components are governed by anisotropy coefficients, enabling controlled, direction-dependent information degradation. This framework provides the theoretical foundation for a novel anisotropic score-based generative model.
    By retaining geometric structure for longer time scales, the backward generative process can exploit residual geometric cues, leading to improved reconstruction fidelity.
    We empirically validate this improvement in a proof-of-concept implementation on unconditional image generation, showing that anisotropic diffusion can achieve superior image quality metrics. We demonstrate consistent improvements in both pixel and latent space experiments over the \gls{sde}-driven baseline as well as over the state-of-the-art Flow Matching approach. Finally, we demonstrate the effectiveness of the introduced anisotropy in a conditional stroke-to-image generation task.
\end{abstract}

\section{Introduction}\label{sec:introduction}

Diffusion-based generative models \citep{song2019generative, ho2020denoising} have become a central paradigm for high-quality image synthesis in both conditional and unconditional settings. Their success is largely due to their ability to model complex high-dimensional data distributions through a gradual destruction process and a learned generative reversal. Beyond image synthesis, this principle has been adopted in a wide range of domains, including video synthesis \citep{xing2024survey}, speech and audio synthesis \citep{kong2020diffwave, huang2022fastdiff}, medical imaging \citep{song2021solving, kazerouni2023diffusion}, and molecular design \citep{weiss2023guided, schneuing2024structure}.

A common feature of many diffusion-based generative models is that the forward process is chosen to progressively remove information from the data in a simple and analytically tractable way. In \gls{sbgm}, this is typically achieved by perturbing the data with noise and learning the score of the resulting perturbed distributions \citep{song2019generative, ho2020denoising, song2021scorebased}. The learned score is then used to reverse the destruction process and generate new samples. While this formulation has proven highly effective, standard forward processes are often isotropic or spatially homogeneous. As a result, they do not explicitly account for geometric structures in the data, such as edges, contours, or coherent regions in images. These methods rely on content-unaware transformations and consequently can struggle to reconstruct geometric structures, leading to artifacts such as truncated or distorted faces and buildings.

This raises a natural question: can the forward destruction process itself be designed to respect such geometric structures? Classical anisotropic diffusion processes suggest that this may be beneficial. In image processing, anisotropic diffusion is used to smooth homogeneous regions while preserving important geometric features such as edges \citep{perona1990scale}. Recent work has also explored related ideas in the context of generative modeling \citep{yu2023constructing, vandersanden2026edge}, indicating that structure-aware perturbations can improve the behavior of diffusion-based models. However, existing models remain limited in their ability to model genuinely anisotropic, state-dependent diffusion dynamics.

In this work, we develop a \gls{sbgm} framework based on anisotropic \acrfullpl{spde}. Our goal is to replace the usual structure-agnostic destruction process by a nonlinear anisotropic diffusion process whose drift and diffusion terms depend on the current state. This allows the forward process to adapt locally to the image geometry, preserving structural information for longer while still driving the data toward a tractable terminal distribution. The resulting formulation provides a mathematically explicit way to incorporate geometry-aware diffusion into \gls{sbgm}.
Our main contributions are the following:
\begin{itemize}
    \item\textbf{Anisotropic diffusion framework}: We introduce an\\ anisotropic diffusion framework for \gls{sbgm} in which the forward process is guided by geometric image structure. The proposed process performs data destruction in a structure-aware manner and is designed to preserve geometric features over longer time scales than isotropic baselines (\cref{sec:diffusion-framework}).
    \item\textbf{Empirical validation}: We demonstrate the practical impact of the framework through proof-of-concept experiments for unconditional image generation in both pixel and latent space (\cref{sec:experiments}). Our results show consistent improvements over isotropic baselines, together with even faster training and inference in our implementation despite the increased theoretical complexity. We further evaluate the proposed anisotropy in a conditional stroke-to-image generation experiment, where it improves reconstruction quality over the corresponding isotropic baseline (\cref{sec:sdedit}).
\end{itemize}

Before presenting our framework, we first review related work and briefly recap the basic principles underlying \glspl{sbgm} in \Cref{sec:related-work,sec:sbgm}.

\section{Related Work}\label{sec:related-work}

\paragraph*{Conceptually related models}

\citet{song2021scorebased} first used time \glspl{sde} for \glspl{sbgm} and showed how existing diffusion models can be unified by an \gls{sde} framework. However, they only considered linear \glspl{sde} with spatially independent diffusion coefficients.

\citet{rissanen2023heat} considered a stochastic heat equation with isotropic noise, which is effectively destroying the data by blurring up to complete dissipation. This is in contrast to earlier approaches that typically destroy data into pure noise. \citet{hoogeboom2022blurring} extended this idea by introducing a temporally increasing isotropic noise term, further refining the blurring process over time.


\paragraph*{State-of-the-art models}

\citet{lipman2022flow} propose \gls{fm}, a simulation-free training method for continuous normalizing flows (CNFs) that regresses vector fields along predefined probability paths. \gls{fm} enables training CNFs with more efficient paths such as optimal transport interpolations, yielding faster sampling and superior sample quality.
We refer to \cref{sec:anisotropic-flow-matching} of the supplemental for a discussion on introducing anisotropy in \gls{fm}.

\citet{zhou2023ddbm} introduce \gls{ddbm}, which generalize diffusion models to map between arbitrary endpoint distributions using learned diffusion bridges. This framework unifies generative modeling paradigms and enables tasks like image translation, achieving strong performance while remaining competitive with state-of-the-art models in standard generation settings. Conditional generative modeling through anisotropic diffusion bridges is part of future work \textemdash\ we elaborate on that in \cref{sec:anisotropic-diffusion-brdige} of the supplemental material.



\paragraph*{Anisotropic models}

Several studies \citep{voleti2022score, yu2023constructing, vandersanden2026edge} explored the role of anisotropic noise in diffusion models. \citet{vandersanden2026edge} propose a structure-aware anisotropy that preserves edges for longer, improving sample quality, particularly in shape-oriented generative tasks.

Our approach shares similarities with this work, as the anisotropic \gls{spde} we introduce is likewise guided by structural image content in both the drift and diffusion terms. However, our method models a genuinely anisotropic and nonlinear diffusion process in which both the drift and diffusion coefficients evolve dynamically with the current state, rather than being determined solely by the initial state. In \citet{vandersanden2026edge}, the edge structure is fixed at time $\timepoint=0$, whereas in our framework the anisotropy is time-adaptive, with the dynamics continuously depending on the edges of the currently transformed state. 

\paragraph*{SPDE-based models}

\citet{lim2023score, lim2024score} also consider generative modeling using \glspl{spde}. The parabolic \gls{spde} studied in \citet{lim2024score} is restricted to spatially-independent diffusion coefficients. This limitation prevents the modeling of spatially varying or anisotropic effects in the forward process, and therefore represents a more constrained setting than our framework that allows for general anisotropic diffusion.


\section{\Acrlong{sbgm}}\label{sec:sbgm}

Generative modeling operates as a two-pass procedure: In the first (\emph{forward}) pass, the information in the data is systematically destroyed. The destruction process is described by a transformation process $\left(\forwardprocess_\timepoint\right)_{\timepoint\in\timedomain}$ evolving over a time domain $\timedomain$, which is either discrete, $\timedomain=\{0,\ldots,\timepointmax\}$, or continuous, $\timedomain=[0,\timepointmax]$, for some \emph{terminal time} $\timepointmax\in[0,\infty)$.
During this pass, the score (i.e. the gradient of the log-density) of the transformation process $\left(\forwardprocess_\timepoint\right)_{\timepoint\in\timedomain}$ is learned by a neural network.
In the \emph{backward} pass, the destroyed data is stochastically reconstructed, resulting in new, previously unseen samples resembling the original data. Because of this dual perspective, the transformation $\left(\forwardprocess_\timepoint\right)_{\timepoint\in\timedomain}$ is called the \emph{forward process}, while its time reversal \begin{equation}\label{eq:backward-process}
    \overline\forwardprocess_\timepoint:=\forwardprocess_{\timepointmax-\timepoint}\eqfor\timepoint\in\timedomain
\end{equation} is referred to as the \setword{\emph{backward process}}{inline:backward-process}.

\subsection{Forward pass (data destruction)}\label{sec:forward-pass}

Generally, in generative modeling, the goal is to learn a \setword{data distribution}{inline:data-distribution} $\datadistribution$ and generate new samples that closely resemble the data. Conceptually, the data distribution $\datadistribution$ is a probability measure on $\mathbb R^\discretedomain$, where $\discretedomain$ is a finite index set. In practice, $\datadistribution$ is unknown and only implicitly represented through a dataset, which we treat as an independent and identically $\datadistribution$-distributed sequence.

Specifically, in \emph{score-based} generative modeling, the data is transformed stochastically into a progressively simpler representation, while the \emph{score} (i.e., the gradient of the log-density) of the resulting forward process is learned. Conceptually, the information contained in the data is gradually destroyed until reaching a sufficiently simple terminal representation, after which the objective is to generate new data resembling the original distribution.

From an analytical perspective, the data is smoothed over time, thereby progressively simplifying the learning task \textemdash\ a principle commonly referred to as \emph{regularization by noise}.

\subsection{Learning the score function}\label{sec:score-estimation}

The score of the forward process provides the necessary information for reconstructing samples in the generative sampling process. Without a score, the reconstruction would be purely deterministic, meaning that the generative sampling process would have to exactly invert the forward
transform.
However, exact inversion is often impossible, as the forward process is designed to progressively degrade information in a way that cannot be deterministically undone.

By incorporating stochasticity into the forward dynamics, we ensure that the degradation is probabilistic, rendering the reverse process well-posed in a statistical sense. We approximate the score using a neural network trained during the forward pass (\cref{sec:forward-pass}) and leverage it to guide
generative sampling as described in \cref{sec:backward-pass}.
That is, for \gls{sbgm}, we need to make sure that the score of the forward process actually exists. Consequently, we assume that $\forwardprocess_\timepoint$ has a positive differentiable \setword{density}{inline:density} $\forwarddensity_\timepoint$ with respect to the $\discretedomain$-dimensional Lebesgue measure for all $\timepoint\in\timedomain$. The \setword{\emph{score}}{inline:score} of the transformation at time $\timepoint\in\timedomain$ is now defined to be \begin{equation}
    \score\left(\timepoint,\;\cdot\;\right):=\nabla\ln\forwarddensity_\timepoint.
\end{equation}

The goal of \gls{sbgm} is now to train a score-based model to find an approximation of $\score$. To this end, we require a suitable metric for measuring the discrepancy between a given approximation $\tilde\score$ and the true score $\score$. This is typically achieved using an $L^2$-norm with respect to a suitable measure \citep[Section~3.3]{song2021scorebased}.

\subsection{Prior sampling}\label{sec:prior-sampling}

The distribution of the forward process at the terminal time $\timepointmax$, namely $\forwardprocess_\timepointmax$, is referred to as the \emph{prior} distribution of the backward (i.e., data generation) process. Formally, a sample from this prior distribution is obtained by initializing the forward process $\left(\forwardprocess_\timepoint\right)_{\timepoint\in\timedomain}$ with a random sample from the data distribution $\datadistribution$ and simulating the process up to the terminal time $\timepointmax$. 

Classically, the forward process is designed such that the prior admits a simple closed-form distribution from which direct sampling is possible. More general forward processes, however, may induce priors without tractable closed-form sampling procedures, requiring the prior samples to be generated through simulation of the forward process itself, as done by \citet{rissanen2023heat}.

\subsection{Backward pass (data generation)}\label{sec:backward-pass}

Once we have trained a score-based model to approximate the score, we can generate new, previously unseen data that resembles the training data distribution $\datadistribution$.
This generation process follows an iterative sampling scheme. It begins by initializing the forward process with a draw from the prior distribution. At each iteration, the scheme applies an optional \emph{predictor} step, followed by an optional \emph{corrector} step \citep[Section~4.2]{song2021scorebased}.

\begin{itemize}[$\circ$]
    \item The predictor step, if applied, propagates the sample backward in time by simulating a step of the backward process.
    \item The corrector step, if applied, refines the sample using the \gls{ula} \citep{roberts1996ula}, treating the sample from the previous iteration as the initial state and targeting the distribution at the corresponding time step. This is feasible because \gls{ula} only requires the gradient of the log-density, which can be estimated using the learned score. However, using the \emph{Metropolis-adjusted} \gls{ula} would necessitate additional density estimation techniques.
\end{itemize}

How sampling in the predictor step can be performed depends entirely on the complexity of the forward process. If the forward process satisfies a Markov property or is given as the solution of an S(P)DE, as is the case for the framework introduced in \cref{sec:diffusion-framework}, specialized sampling techniques become available. We refer to \cref{sec:backward-sampling} of the supplemental material for details.

\section{Anisotropic diffusion framework}\label{sec:diffusion-framework}

\paragraph*{Motivation}

The performance of a score-based generative model, as measured by the visual quality of the generated data samples and by metrics quantifying their resemblance to the underlying data distribution, depends (aside from the chosen network architecture) primarily on the design of the forward process.

Existing approaches are mostly based on \emph{isotropic} transformations. However,
isotropic methods do not explicitly encode local geometric structure in the forward process, which can hinder the reconstruction of geometric features during generative sampling. Achieving geometry-aware transformations instead requires nonlinear, \emph{anisotropic} dynamics.
Existing anisotropic approaches either employ anisotropic noise that is not state-dependent \citep{yu2023constructing}, and therefore likewise cannot adapt to or capture the underlying geometry, or they preserve geometric information only at the initial time point of the transformation and do not adapt to geometric structures as they evolve throughout the transformation process \citep{vandersanden2026edge}.

To enable genuinely anisotropic transformations, we extend the \gls{sde}-based approach of \citet{song2021scorebased} by formulating the forward process as the solution to a stochastic \emph{partial} differential equation. This formulation naturally incorporates spatial derivatives, enabling structured and anisotropic transformations of images.

For this purpose, when formulating the forward process as an \gls{spde}, we treat each image channel as a function on a continuous rectangular domain rather than as a discrete array of pixel values.

\subsection{Anisotropic forward process: The theoretical model}

We now specify the \gls{spde} defining our forward process. The process progressively destroys information in a structure-aware, anisotropic manner while preserving geometric cues over longer time scales, thereby providing the generative backward process with richer structural information for reconstruction.
To this end, we model the forward process $\left(\forwardprocess_\timepoint\right)_{\timepoint\in\timedomain}$ as the \emph{formal} solution of \begin{equation}\label{eq:forward-equation}
    \dd{\forwardprocess_\timepoint}=\drift\left(\timepoint,\forwardprocess_\timepoint\right)\dd{\timepoint}+\diffusioncoefficient\left(\timepoint,\forwardprocess_\timepoint\right)\dd{\wienerprocess_\timepoint},
\end{equation} where \begin{equation}\label{eq:forward-drift}
    \drift(\timepoint,\evolutionpoint):=\nabla\cdot\anisotropycoefficient_1(\timepoint,\nabla\evolutionpoint)\nabla\evolutionpoint
\end{equation} and
\begin{equation}\label{eq:forward-diffusion-coefficient}
    \diffusioncoefficient(\timepoint,\evolutionpoint)\noisemode:=\anisotropycoefficient_2(\timepoint,\nabla\evolutionpoint)\noisemode.
\end{equation} The coefficients $\anisotropycoefficient_\indexsetelement$ are chosen as
\begin{equation}
    \anisotropycoefficient_\indexsetelement(\timepoint,\point):=\frac{\diffusivitytransition_\indexsetelement(\timepoint)}{\displaystyle\sqrt{1+\left\|\frac\point{\anisotropytransition_\indexsetelement(\timepoint)}\right\|^2}},
\end{equation} where $\diffusivitytransition_\indexsetelement$ and $\anisotropytransition_\indexsetelement$ are functions controlling the intensity of destruction and strength of geometry preservation. The driving noise process $\left(\wienerprocess_\timepoint\right)_{\timepoint\in\timedomain}$ is a cylindrical Wiener process \citep{daprato2014evolution}.

The \gls{spde} in \autoref{eq:forward-equation} is a natural stochastic extension of the classical Perona--Malik diffusion \citep{perona1990scale}. Intuitively, diffusion is reduced near strong gradients, allowing geometric structures such as edges to persist over longer time scales.

\paragraph*{Design intuition}

The \emph{drift} $\drift$ introduces deterministic smoothing that respects the local image geometry through the \emph{anisotropy coefficient} $\anisotropytransition_1$. The overall smoothing strength is controlled by the \emph{diffusivity coefficient} $\diffusivitytransition_1$, while the local gradients $\nabla\forwardprocess_\timepoint$ determine the smoothing direction and magnitude.
The diffusion coefficient $\diffusioncoefficient$ injects stochastic noise into the image. Its overall intensity is controlled by $\diffusivitytransition_2$, while the anisotropy coefficient $\anisotropytransition_2$ modulates the noise in a structure-aware manner.

\begin{figure}[t]
    \centering
    \setlength{\tabcolsep}{1pt}
    \begin{tabular}{ccc}
        \includegraphics[width=.14\linewidth]{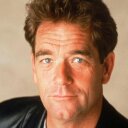}
        &
        \includegraphics[width=.14\linewidth]{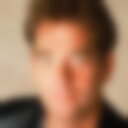}
        &
        \includegraphics[width=.14\linewidth]{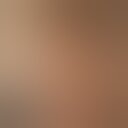}
        \\[-.1cm]
        $\diffusivitytransition_1=0$
        &
        $\diffusivitytransition_1=1$
        &
        $\diffusivitytransition_1=0.5\to1000$
        \\
        \includegraphics[width=.14\linewidth]{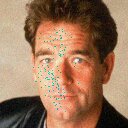}
        &
        \includegraphics[width=.14\linewidth]{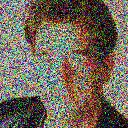}
        &
        \includegraphics[width=.14\linewidth]{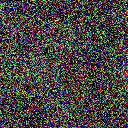}
        \\[-.1cm]
        $\diffusivitytransition_2=1\mathrm{e{-}2}$
        &
        $\diffusivitytransition_2=1\mathrm{e{-}1}$
        &
        $\diffusivitytransition_2=1$
        \\
        \includegraphics[width=.14\linewidth]{figures/diffusion/diffusivity_coefficient/diffusion_diffusivity_coefficient_2.jpg}
        &
        \includegraphics[width=.14\linewidth]{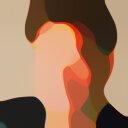}
        &
        \includegraphics[width=.14\linewidth]{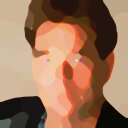}
        \\[-.1cm]
        $\anisotropytransition_1=\infty$
        &
        $\anisotropytransition_1=1\mathrm{e{-}2}$
        &
        $\anisotropytransition_1=1\mathrm{e{-}3}$
        \\
        \includegraphics[width=.14\linewidth]{figures/diffusion/intensity_coefficient/diffusion_intensity_coefficient_3.jpg}
        &
        \includegraphics[width=.14\linewidth]{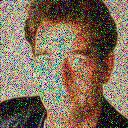}
        &
        \includegraphics[width=.14\linewidth]{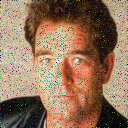}
        \\[-.1cm]
        $\anisotropytransition_2=\infty$
        &
        $\anisotropytransition_2=1\mathrm{e{-}2}$
        &
        $\anisotropytransition_2=1\mathrm{e{-}3}$
    \end{tabular}
    \caption{
        Effect of the main components of our anisotropic diffusion framework on an image with resolution $128\times128$. From top to bottom, the rows illustrate:
        (1) the impact of the diffusivity coefficient $\diffusivitytransition_1$, where the third column uses a geometric transition of the form $\diffusivitytransition_1(\timepoint)=\diffusivitytransition_1^{\textnormal{min}}\left(\diffusivitytransition_1^{\textnormal{max}}/\diffusivitytransition_1^{\textnormal{min}}\right)^{\timepoint/\timepointmax}$ with $\diffusivitytransition_1^{\textnormal{min}}=0.5$ and $\diffusivitytransition_1^{\textnormal{max}}=1k$, and where the effect of the diffusion coefficient $\diffusioncoefficient$ was disabled by setting $\diffusivitytransition_2=0$;
        (2) the impact of the intensity coefficient $\diffusivitytransition_2$, where the effect of the drift $\drift$ was disabled by setting $\diffusivitytransition_1=0$;
        (3) the impact of the anisotropy coefficient $\anisotropytransition_1$, where the effect of the diffusion coefficient $\diffusioncoefficient$ was disabled by setting $\diffusivitytransition_2=0$; and
        (4) the impact of the anisotropy coefficient $\anisotropytransition_2$, where the effect of the drift $\drift$ was disabled by setting $\diffusivitytransition_1=0$.
    }
    \label{fig:design}
\end{figure}

Together, the drift and diffusion coefficient provide complementary mechanisms for destroying information: the drift performs deterministic smoothing, whereas the diffusion coefficient introduces stochastic perturbations.

To enable generative reconstruction, the forward process must destroy information sufficiently over time. Since anisotropy preserves geometric information to some extent, the process must eventually transition toward stronger information destruction. This can, for example, be achieved by increasing the stochastic noise intensity at later time points or by gradually transitioning the drift toward isotropic diffusion. The latter corresponds to choosing $\anisotropytransition_1(\timepoint)\to\infty$ as $\timepoint\to\timepointmax$.
We illustrate the effect of the individual components of the proposed framework in \cref{fig:design}. Additional details are provided in the supplemental material.

\subsection{Practical forward and backward processes}

From a strictly mathematical perspective, ensuring correctness of the generative modeling methodology requires that the backward process is the \emph{exact} time-reversal of the forward process and that both forward and backward processes are \emph{exactly} simulatable. However, for a complex \gls{spde} like \cref{eq:forward-equation}, exact simulation is impossible. Both spatial and temporal discretization must be performed to obtain a practically simulatable process.

Spatial discretization \textemdash\ whether via Galerkin methods \citep{lim2024score} or finite differences as in our numerical scheme \textemdash\ inevitably
leads to a projection onto a finite-dimensional \gls{sde} of the form
\begin{equation}\label{eq:practical-forward-process}
    \dd{\tilde\forwardprocess_\timepoint}=\tilde\drift\left(\timepoint,\tilde\forwardprocess_\timepoint\right)\dd{\timepoint}+\tilde\diffusioncoefficient\left(\timepoint,\tilde\forwardprocess_\timepoint\right)\dd{\tilde\wienerprocess_\timepoint}\eqforall\timepoint\in\timedomain,
\end{equation}
where $\tilde\drift:\mathbb R^\discretedomain\to\mathbb R^\discretedomain$, $\tilde\diffusioncoefficient:\mathbb R^\discretedomain\to\mathbb R^{\discretedomain\times\discretedomain}$, and $(\tilde\wienerprocess_\timepoint)_{\timepoint\ge0}$ is a $\discretedomain$-dimensional Brownian motion for some $\discretedomain\in\mathbb N$.

This procedure unavoidably introduces approximation error, which is further compounded by the subsequent temporal discretization. In special cases \textemdash\ such as the parabolic \glspl{spde} with additive noise considered in \citet{lim2024score} \textemdash\ this may be acceptable. Nevertheless, if one uses the time-reversal of the infinite-dimensional \gls{spde} as the backward process, a mismatch arises between the processes that can actually be simulated and the theoretical foundation of their use, and exact correctness of the backward process is no longer guaranteed.

In our framework, we consider a substantially more complex \gls{spde} \textemdash\ with gradient-dependent nonlinear drift and gradient-dependent multiplicative noise \textemdash\ than in previous \gls{spde}-based \glspl{sbgm} (e.g., \citet{lim2024score}). In our case, no fixed eigenbasis diagonalizes the drift operator or the noise, and modal decoupling is unavailable.

We therefore adopt a different strategy: we first apply spatial discretization, thereby defining the actual forward process directly as the finite-dimensional \gls{sde} \cref{eq:practical-forward-process}. The backward process is then taken as the time-reversal of this \gls{sde}, ensuring that both forward and backward dynamics are defined at the same level of approximation. As a result, the only simulation error arises from temporal discretization, rather than from a combination of spatial \emph{and} temporal discretizations.
The concrete numerical scheme leading to the \gls{sde} \cref{eq:practical-forward-process} used in our experiments in \cref{sec:experiments} is described in \cref{sec:numerical-simulation} of the supplemental. Theoretical existence of a solution to \cref{eq:practical-forward-process} is verified by \citet[Theorem~9.11]{pascucci2011option}. That the corresponding backward process also satisfies an \gls{sde} follows from classical results \citep{Haussmann1986diffusion, anderson1982reversal}. For the explicit form of this \gls{sde} we refer to \cref{sec:backward-sampling} of the supplemental.



\subsection{Residual dependence on the initial state}\label{sec:residual-dependence}

Introducing anisotropy inherently induces a residual dependence on the initial state. Our framework contains many user-definable parameters.
For meaningful use in generative modeling, these parameters should be chosen such that the information contained in the initial state is almost entirely degraded by the terminal time $\timepointmax$.
Conceptually, anisotropy should only serve to prolong the preservation of certain structures (e.g., edges)
to facilitate the reconstruction of geometric features during data generation.

For specific parameter choices in our framework, the forward process reduces to an Ornstein–Uhlenbeck process (as in the instance described in \cref{sec:exp_stoch_heat_iso_noise}), or it can be designed such that the distribution at the terminal time $\timepointmax$ is close to a known Gaussian distribution (as in the instance described in \cref{sec:exp_process_aniso_heat_iso_noise}). In such cases, prior sampling can be performed by directly sampling from the closed-form of the prior distribution.

Whether the prior distribution admits a closed form (or a tractable approximation) depends on the chosen parameters. If not, a sample from the prior can be generated by simulating the forward process up to the terminal time. In either case,
the construction remains conceptually well-defined;
see our discussion
in \cref{sec:prior-sampling}.

\section{Numerical study}\label{sec:experiments}

Intuitively, the anisotropy introduced by our anisotropic diffusion framework helps preserve structural information over longer time scales during the forward
transform.
This, in turn, facilitates learning and reconstruction of geometric features during data generation.

We empirically validate this intuition through a numerical study on unconditional image generation. We compare our framework against four baselines: \citet{rissanen2023heat}; the \emph{variance exploding SDE} (\textsc{VESDE}) from \citet{song2021scorebased}; the \emph{Flow Matching \slash\ Optimal Transport} method of \citet{lipman2022flow}, which is
among the state-of-the-art baselines at the time of writing; and the anisotropic-noise approach of \citet{vandersanden2026edge}, whose anisotropy is fixed from the initial state rather than evolving with the current state.

We focus on two specific instances of our framework: an isotropic version (\emph{Ours (isotropic)} described in \cref{sec:exp_stoch_heat_iso_noise}) and an anisotropic version (\emph{Ours (anisotropic)} described in \cref{sec:exp_process_aniso_heat_iso_noise}) of a stochastic heat equation. The isotropic variant is included primarily for educational purposes, as it demonstrates that the model equation of \citet{rissanen2023heat} arises as a special case of our framework.

Both \citet{rissanen2023heat} and \citet{song2021scorebased} provide especially relevant baselines, since they are likewise based on S(P)DEs. In contrast to our approach, however, their drift and diffusion coefficients are isotropic. Conceptually, the only difference between their methods and \emph{Ours (anisotropic)} is the introduction of anisotropy.
This comparison is designed to isolate the effect of anisotropy as much as possible within our implementation.
%
%
%
%
%
We now describe the specific instances of our framework considered in the numerical study (\cref{sec:song,sec:exp_process_aniso_heat_iso_noise,sec:exp_stoch_heat_iso_noise}), followed by a detailed account of the experiments in \cref{sec:results}.

\subsection{Pure isotropic noise (\citet{song2021scorebased})}\label{sec:song}

\begin{equation}\label{eq:ve-sde}
    \dd{\forwardprocess_\timepoint}=\diffusivitytransition_2(\timepoint)\dd{\wienerprocess_\timepoint}
\end{equation}

The \gls{spde} formulated in \autoref{eq:ve-sde} defines a Gaussian process, of which the \textsc{VESDE} considered in \citet{song2021scorebased} is a specific instance. It has a vanishing drift term, $\drift = 0$, and an isotropic diffusion coefficient, $\diffusioncoefficient$. Intuitively, this corresponds to a process in which an increasing amount of noise is added to the data over time. 




\subsection{Anisotropic stochastic heat equation\\ with isotropic noise (Ours (anisotropic))}\label{sec:exp_process_aniso_heat_iso_noise}

\begin{equation}\label{eq:anisotropic-heat-equation}
    \dd{\forwardprocess_\timepoint}=\nabla\cdot\frac{\diffusivitytransition_1(\timepoint)}{\displaystyle\sqrt{1+\left\|\frac{\nabla\forwardprocess_\timepoint}{\anisotropytransition_1}\right\|^2}}\nabla\forwardprocess_\timepoint\dd{\timepoint}+\diffusivitytransition_2(\timepoint)\dd{\wienerprocess_\timepoint}
\end{equation}

The \gls{spde} in \autoref{eq:anisotropic-heat-equation} is a genuinely anisotropic instance of our anisotropic diffusion framework \cref{eq:forward-equation}, where the drift gradually transitions from anisotropy to isotropy while the diffusion coefficient remains isotropic.  
We consider \emph{geometric} transitions of the form  
\begin{equation}\label{eq:geometric-transition}
    \diffusivitytransition_\sequenceindex(\timepoint):=\diffusivitytransition^{\textnormal{min}}_\sequenceindex\left(\frac{\diffusivitytransition^{\textnormal{max}}_\sequenceindex}{\diffusivitytransition^{\textnormal{min}}_\sequenceindex}\right)^{\frac\timepoint\timepointmax}
\end{equation}  
for $0<\diffusivitytransition^{\textnormal{min}}_\sequenceindex<\diffusivitytransition^{\textnormal{max}}_\sequenceindex$ (other common transitions are shown in \cref{fig:trans_fns_divergence_fns_vis}). Specifically, $\diffusivitytransition_1$ increases geometrically from $\diffusivitytransition^{\textnormal{min}}_1=0.5$ to $\diffusivitytransition^{\textnormal{max}}_1=2\cdot\textnormal{image size}$, while the anisotropy coefficient $\anisotropytransition_1$ ensures a slow transition from anisotropy to isotropy via  
\begin{equation}
    \anisotropytransition_1(\timepoint):=\anisotropytransition^{\textnormal{min}}_1\frac{e^{\exponentialparam\timepointmax}-1}{e^{\exponentialparam\left(\timepointmax-\timepoint\right)}-1}
\end{equation}  
with $\anisotropytransition^{\textnormal{min}}_1=0.025$ and $\exponentialparam=1/2$ (see \cref{fig:trans_fns_divergence_fns_vis} for a visualization).  
The intensity coefficient $\diffusivitytransition_2$ also increases geometrically, from $\diffusivitytransition^{\textnormal{min}}_2=0.01$ to $\diffusivitytransition^{\textnormal{max}}_2=2.0$. With $\anisotropytransition_2\equiv\infty$, the diffusion coefficient remains spatially isotropic throughout.  
We set the noise correlation length to $\wienercovariancecorrelationlength=0$, leading to a \emph{cylindrical} Wiener process~$(\wienerprocess_\timepoint)_{\timepoint\in\timedomain}$ and hence spatially white noise \citep{daprato2014evolution}. The corresponding forward and backward processes are visualized in \cref{fig:anisotropic-forward-and-backward-sampling} in the supplemental material.

Since $\anisotropytransition_1(\timepoint)\to\infty$ as $\timepoint\to\timepointmax$, the numerical simulation \cref{eq:practical-forward-process} of the \gls{spde} \cref{eq:anisotropic-heat-equation} is, at least approximately, conditionally Gaussian given the initial state. Hence, prior sampling can be performed from a closed-form (Gaussian) distribution in the implementation.


\subsection{Stochastic heat equation with isotropic noise\\ (\citet{rissanen2023heat} and Ours (isotropic))}\label{sec:exp_stoch_heat_iso_noise}

\begin{equation}\label{eq:stochastic-heat-with-additive-noise-repeated}
    \dd{\forwardprocess_\timepoint}=\diffusivitytransition_1(\timepoint)\Delta\forwardprocess_\timepoint\dd{\timepoint}+\diffusivitytransition_2(\timepoint)\dd{\wienerprocess_\timepoint}
\end{equation}

The \gls{spde} in \autoref{eq:stochastic-heat-with-additive-noise-repeated} has an isotropic drift $\drift$ and a \emph{small-scale} isotropic diffusion coefficient $\diffusioncoefficient$. It closely resembles the forward process considered by \citet{rissanen2023heat}. Intuitively, the data is smoothed over time, and a small amount of noise is injected to make the forward process stochastic, which is required to ensure that the reverse process is well-posed (see \cref{sec:score-estimation}).  

For $\diffusivitytransition_1$, we again use a geometric transition \cref{eq:geometric-transition} with $\diffusivitytransition^{\textnormal{min}}_1=0.5$ and $\diffusivitytransition^{\textnormal{max}}_1=2\cdot\textnormal{image size}$. A minor difference from \citet{rissanen2023heat} is that we do not keep the intensity coefficient $\diffusivitytransition_2$ constant; instead, it increases slightly over time (while remaining small) under a geometric transition \eqref{eq:geometric-transition} with $\diffusivitytransition^{\textnormal{min}}_2=0.01$ and $\diffusivitytransition^{\textnormal{max}}_2=0.5$. Formally, the anisotropy coefficients are fixed as $\anisotropytransition_1=\anisotropytransition_2\equiv\infty$. The noise correlation length is again set to $\wienercovariancecorrelationlength=0$, so we also work with a cylindrical Wiener process~$(\wienerprocess_\timepoint)_{\timepoint\in\timedomain}$ here.  

With the parameter choices described above, the numerical simulation \cref{eq:practical-forward-process} of the \gls{spde} \cref{eq:stochastic-heat-with-additive-noise-repeated} is conditionally Gaussian given the initial state. Consequently, prior sampling can be carried out from a closed-form (Gaussian) distribution in the implementation.



\subsection{Pixel Space Experiments}\label{sec:results}

In our experiments, our methods \emph{Ours (anisotropic)} and \emph{Ours (isotropic)} used both the predictor and corrector steps described in \cref{sec:backward-pass}. Each corrector step consists of a single \gls{ula} iteration.

\paragraph*{Test datasets}

We trained all generative models on\newline \textsc{Cifar10}~\citep{krizhevsky2009learning}, \textsc{CelebA}~\citep{conficcvLiuLWT15}, \textsc{ImageNet2012}~\citep{ILSVRC15}, \textsc{LSUN\slash bedroom}~\citep{yu2015lsun}, and \textsc{LSUN\slash church\_outdoor}~\citep{yu2015lsun}.

\paragraph*{Evaluation}\label{sec:evaluation}


We assess the quality of generated samples using standard metrics: Inception Score (IS) \citep{NIPS2016_8a3363ab}, Fréchet Inception Distance (FID) \citep{NIPS2017_8a1d6947}, and Kernel Inception Distance (KID) \citep{bińkowski2021demystifyingmmdgans}. To ensure consistent evaluation, we used the implementation \citep{song2021github} provided by \citet{song2021scorebased} and regenerated all samples \textemdash\ including those for baseline methods \textemdash\ to compute these metrics. Because metric implementations vary slightly across toolsets, our reported values may differ from those originally published. This makes it particularly important that all models are evaluated under identical conditions.

\paragraph*{Quantitative comparison}\label{sec:comparison}

In \cref{tab:metrics}, we report the evaluation metrics of samples generated by the different methods on the test datasets.
All values reported for \emph{Ours (isotropic)} and \emph{Ours (anisotropic)} are based on training from scratch (i.e., without initialization from a pre-trained checkpoint) for 200{,}000 steps. Since no checkpoints are provided by the authors, we retrained the method of \citet{lipman2022flow} on all datasets using the official configuration from their codebase for 200{,}000 steps as well. The same procedure was applied to \citet{rissanen2023heat}, although we restricted evaluation to \textsc{Cifar10}, given the method’s already non-competitive performance on this dataset. For \citet{song2021scorebased}, we trained \textsc{CelebA} and \textsc{ImageNet2012} from scratch, while for \textsc{Cifar10}, \textsc{LSUN\slash bedroom} and \textsc{LSUN\slash church\_outdoor} we relied on the official checkpoints released by the authors \citep{song2021github}.
All evaluation metrics were computed using \citet{stein2023exposingflawsgenerativemodel}.

In \autoref{tab:fidelity-diversity}, we additionally evaluate the fidelity--diversity trade-off across datasets, showing that the observed gains in fidelity are not accompanied by a reduction in diversity.

\paragraph*{Fine-tuning a pretrained model}

\begin{figure}[H]
    \centering
    \begin{minipage}[t]{.48\textwidth}
        \centering
        \begin{tikzpicture}[scale=.85]
            \begin{axis}[
                scale only axis,
                scale = .6,
                xlabel = {Training iterations (in thousands)},
                grid = major,
                legend pos = north west,
                legend cell align = {left},
                xtick = {10, 20, 30, 40, 50, 60, 70, 80, 90, 100},
                xticklabels = {$10$, $20$, $30$, $40$, $50$, $60$, $70$, $80$, $90$, $100$},
                ytick = {9.4, 9.5, 9.6, 9.7, 9.8},
                tick align = outside,
            ]
                \addplot[color = red, thick, mark = *] table[x index=0, y index = 1, col sep = tab] {results/cifar10_ours_anisotropic.error};
                \addlegendentry{IS $\uparrow$}
            \end{axis}
        \end{tikzpicture}
    \end{minipage}%
    \begin{minipage}[t]{.48\textwidth}
        \centering
        \begin{tikzpicture}[scale=.85]
            \begin{axis}[
                scale only axis,
                scale = .6,
                xlabel = {Training iterations (in thousands)},
                grid = major,
                legend pos = north east,
                legend cell align = {left},
                xtick = {10, 20, 30, 40, 50, 60, 70, 80, 90, 100},
                xticklabels = {$10$, $20$, $30$, $40$, $50$, $60$, $70$, $80$, $90$, $100$},
                ytick = {2, 2.5, 3, 3.5, 4, 4.5, 5, 5.5},
                tick align = outside,
            ]
                \addplot[color = blue, thick, mark = *] table[x index=0, y index = 2, col sep = tab] {results/cifar10_ours_anisotropic.error};
                \addlegendentry{FID $\downarrow$}
            \end{axis}
        \end{tikzpicture}
    \end{minipage}
    \caption{
        Evaluation metrics on \textsc{Cifar10} between 10k and 100k training iterations with \emph{Ours (anisotropic)} in a fine-tuning experiment, where training was initialized from the checkpoint of \citet{song2021scorebased}.
        For reference, applying our sampling procedure directly to that checkpoint \textemdash\ without further training \textemdash\ results in an IS of 1, FID of 678.3, and KID of 0.9. After 100k training iterations, these values improve substantially to an IS of 10.0, FID of 2.0, and KID of 6.4e-4
    }
    \label{fig:fine-tuning}
\end{figure}
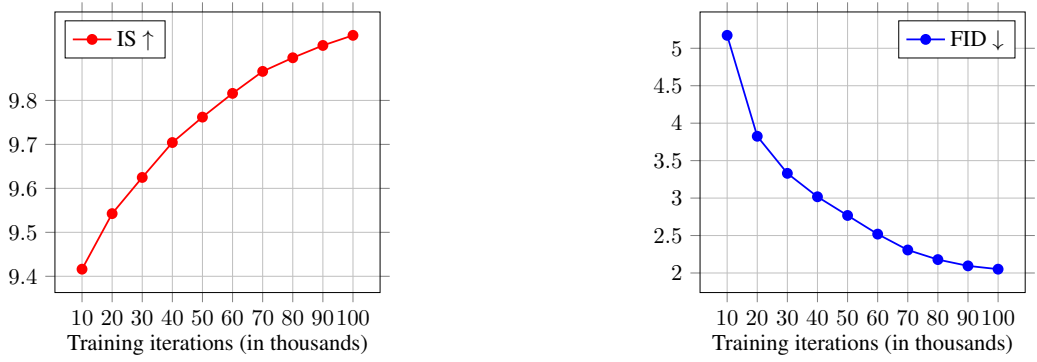

\cref{fig:fine-tuning} illustrates how quickly (i.e., after how few training iterations) \emph{Ours (anisotropic)} improves the evaluation metrics when initialized from a model pretrained with \citet{song2021scorebased}. Notably, according to the authors, continuing training with their own method did not yield
metric improvements.

\paragraph*{Qualitative comparison}

\begin{figure}[H]
    \setlength{\tabcolsep}{.1cm}
    \renewcommand{\arraystretch}{0}
    \begin{tabular}{@{}cc@{}}
        \begin{minipage}{.5\textwidth}
            \setlength{\tabcolsep}{0cm}
            \begin{tabular}{@{}ccc@{}}
                \includegraphics[width = \linewidth/3]{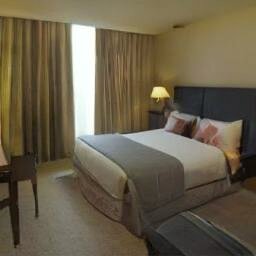}
                &
                \includegraphics[width = \linewidth/3]{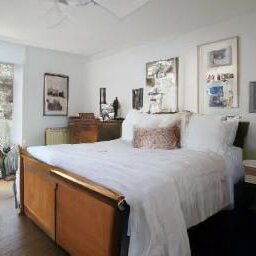}
                &
                \includegraphics[width = \linewidth/3]{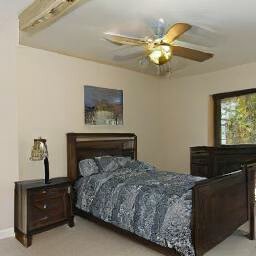}
                \\
                \includegraphics[width = \linewidth/3]{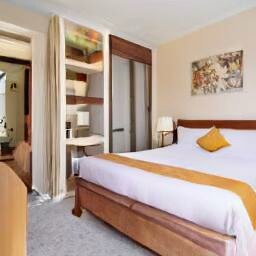}
                &
                \includegraphics[width = \linewidth/3]{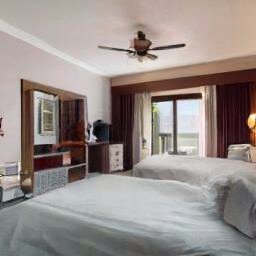}
                &
                \includegraphics[width = \linewidth/3]{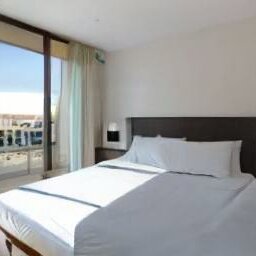}
            \end{tabular}
        \end{minipage}
        &
        \begin{minipage}{.5\textwidth}
            \setlength{\tabcolsep}{0cm}
            \begin{tabular}{@{}ccc@{}}
                \includegraphics[width = \linewidth/3]{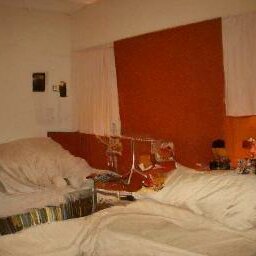} &
                \includegraphics[width = \linewidth/3]{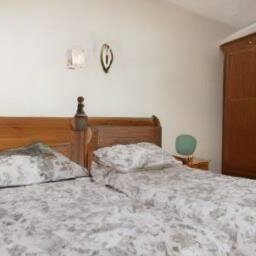} &
                \includegraphics[width = \linewidth/3]{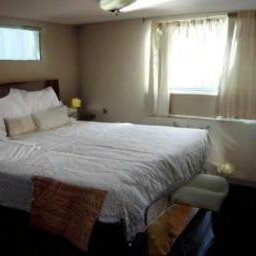}
                \\
                \includegraphics[width = \linewidth/3]{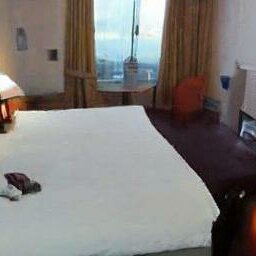}
                &
                \includegraphics[width = \linewidth/3]{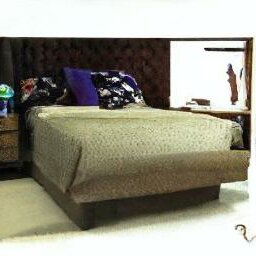}
                &
                \includegraphics[width = \linewidth/3]{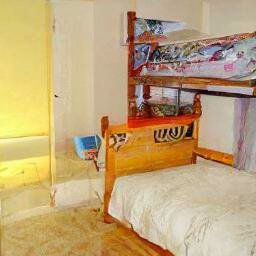}
            \end{tabular}
        \end{minipage}
        \\\noalign{\vskip .1cm}
        \footnotesize \emph{Ours (anisotropic)} (\textbf{FID: 6.1})
        &
        \footnotesize \citet{song2021scorebased} (FID: 14.9)
    \end{tabular}
    \caption{
        Generated samples on \textsc{LSUN\slash bedroom} from our pixel space experiment (\cref{sec:results}). Left: Samples generated by \emph{Ours (anisotropic)} (see \cref{sec:exp_process_aniso_heat_iso_noise}), using a model trained from scratch \textemdash\ i.e., without initialization from a pre-trained checkpoint. Right: Samples generated by \citet{song2021scorebased}, using a model trained from scratch as well. \emph{Ours (anisotropic)} more faithfully resembles the geometric structure of the dataset.
    }
    \label{fig:bedroom}
\end{figure}

In \cref{fig:teaser,fig:bedroom} and \cref{sec:uncurated-samples-cifar10,sec:uncurated-samples-celeba,sec:uncurated-samples-imagenet2012,sec:uncurated-samples-lsun,sec:uncurated-samples-bedroom} of the supplemental material, we present uncurated generated samples on \textsc{Cifar10} (32×32), \textsc{CelebA} (64×64), \textsc{ImageNet2012} (64×64), \textsc{LSUN\slash bedroom} (256×256), and \textsc{LSUN\slash church\_outdoor} (256×256).

\paragraph*{Computational costs}

In \cref{fig:cost-comparison}, we compare computational costs and show that our method achieves lower training and inference times than \citet{song2021scorebased} and \citet{lipman2022flow}.
We simulated the forward and backward processes up to a terminal time of $\timepointmax=2$ using a numerical step size of $\Delta\timepoint=1\mathrm{e-}2$, resulting in 199 score function evaluations in the predictor step during backward-process simulation. Our choice of step size was intentionally conservative to avoid numerical discretization artifacts affecting the evaluation. The number of score evaluations depends directly on the chosen discretization, and in practice larger or adaptive step sizes may be used. Due to the already large hyperparameter space and hardware constraints, we did not perform an ablation study over $\Delta\timepoint$.

\begin{figure}[H]
    \centering
    \begin{subtable}[t]{.48\textwidth}
    \centering
    \caption{Training costs (per processed dataset image)}
    \begin{tabular}{l c}
        \toprule
        \textbf{Model} & \textbf{Seconds per sample} \\
        \midrule
        Ours (anisotropic)         & \textbf{0.285} \\
        \citet{song2021scorebased} & 0.665 \\
        \citet{lipman2022flow}     & 0.293 \\
        \bottomrule
    \end{tabular}
    \end{subtable}
    \\
    \begin{subtable}[t]{.48\textwidth}
    \centering
    \caption{Inference costs (per generated image)}
    \begin{tabular}{l c}
        \toprule
        \textbf{Model} & \textbf{Seconds per sample} \\
        \midrule
        Ours (anisotropic)         & \textbf{0.165} \\
        \citet{song2021scorebased} & 0.572 \\
        \citet{lipman2022flow}     & 0.355 \\
        \bottomrule
    \end{tabular}
    \end{subtable}
    
    \caption{Normalized (a) training and (b) inference costs on \textsc{LSUN\slash church\_outdoor} (256×256), where training cost is measured as the time required for a single loss evaluation, reported per processed dataset sample.
    Despite a theoretically more demanding drift and diffusion coefficient, our implementation achieves significantly improved inference times compared to \citet{song2021scorebased}, due to the numerical scheme described in \autoref{sec:numerical-simulation}.}
    \label{fig:cost-comparison}
\end{figure}

\paragraph*{Hardware resources}\label{sec:hardware}

All experiments were conducted on a server equipped with 8$\times$ NVIDIA Tesla H100 NVL GPUs (94 GB HBM3 each, PCIe 5.0) and 2$\times$ AMD EPYC 9554 CPUs (64 cores / 128 threads each, 3.1–3.75 GHz, Genoa microarchitecture, 256 MB L3 cache).

\paragraph*{Hyperparameters and architecture}\label{par:architecture}

For the implementation of our framework, we adopted the NCSN++ (continuous) network architecture from \citet{song2021scorebased}. The training parameters \textemdash\ in particular, a learning rate of 2e-4 using the Adam optimizer~\citep{kingma2014adam} \textemdash\ are taken from \citet{song2021scorebased} for both \textsc{Cifar10}, \textsc{CelebA} and \textsc{LSUN\slash church\_outdoor}. For \textsc{ImageNet2012}, we reuse the parameter settings for \textsc{CelebA}.

\paragraph*{Limitations}\label{sec:limitations}


The flexibility of our framework opens the door to exploring a wide range of anisotropic forward processes beyond \emph{Ours (anisotropic)}, which may further enhance quality. However, due to hardware resource limitations, this study is limited to the single parameter configuration given by \emph{Ours (anisotropic)} and the restricted set of baseline methods mentioned above.

\subsection{Latent Space Experiment}\label{sec:experiments-latent-space}

We also conducted a latent space experiment using \textit{Ours (anisotropic)} on the \textsc{LSUN\slash church\_outdoor} (256x256) and \textsc{LSUN\slash bedroom} (256x256) datasets \textemdash\ effectively reducing the dimension to 64x64. 
For the latent representation, we employed the pretrained variational autoencoder from \texttt{stabilityai/sd-vae-ft-mse}~\citep{rombach2022latent}.
The generative performance metrics on \textsc{LSUN\slash church\_outdoor} / \textsc{LSUN\slash bedroom} are:

\begin{center}
    \begin{tabular}{lccc}
        \toprule
        & IS $\uparrow$ & FID $\downarrow$ & KID $\downarrow$ \\
        \midrule
        \textbf{Ours (anisotropic)} 
            & \textbf{3.9} / \textbf{3.8}
            & \textbf{3.9} / \textbf{5.7}
            & \textbf{2.3e-3} / \textbf{2.5e-3} \\
        \citet{song2021scorebased}
            & 2.1 / 2.0
            & 13.4 / 14.1
            & 7.5e-3 / 8.1e-3 \\
        \bottomrule
    \end{tabular}
\end{center}

\begin{figure}[H]
    \setlength{\tabcolsep}{.1cm}
    \renewcommand{\arraystretch}{0}
    \begin{tabular}{@{}cc@{}}
        \begin{minipage}{.5\textwidth}
            \setlength{\tabcolsep}{0cm}
            \begin{tabular}{@{}ccc@{}}
                \includegraphics[width = \linewidth/3]{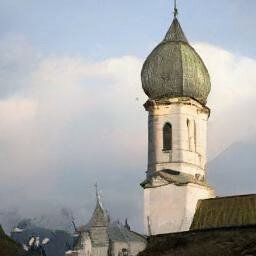}
                &
                \includegraphics[width = \linewidth/3]{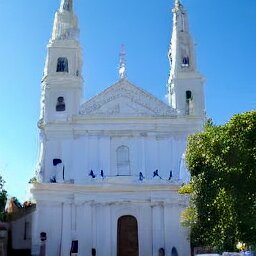}
                &
                \includegraphics[width = \linewidth/3]{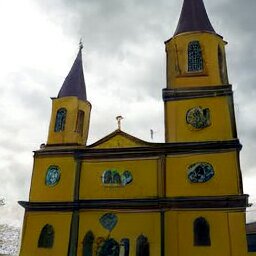}
                \\
                \includegraphics[width = \linewidth/3]{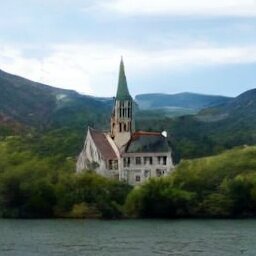}
                &
                \includegraphics[width = \linewidth/3]{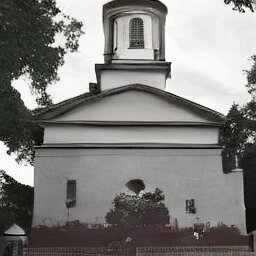}
                &
                \includegraphics[width = \linewidth/3]{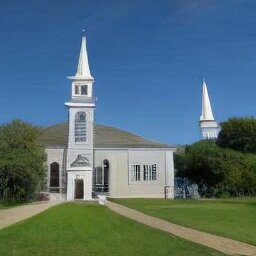}
            \end{tabular}
        \end{minipage}
        &
        \begin{minipage}{.5\textwidth}
            \setlength{\tabcolsep}{0cm}
            \begin{tabular}{@{}ccc@{}}
                \includegraphics[width = \linewidth/3]{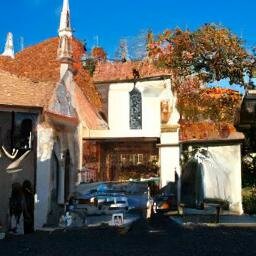} &
                \includegraphics[width = \linewidth/3]{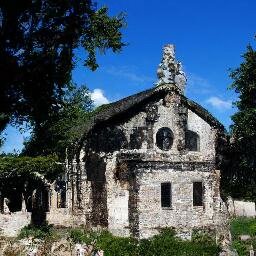} &
                \includegraphics[width = \linewidth/3]{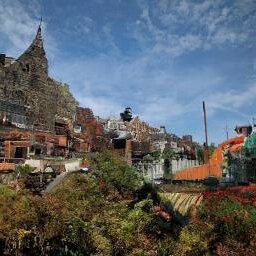}
                \\
                \includegraphics[width = \linewidth/3]{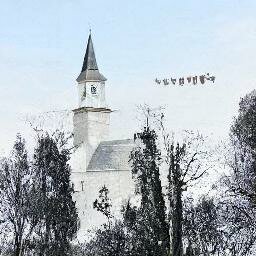}
                &
                \includegraphics[width = \linewidth/3]{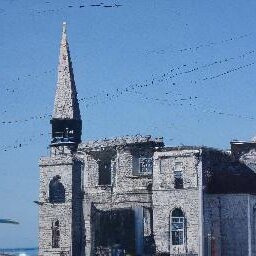}
                &
                \includegraphics[width = \linewidth/3]{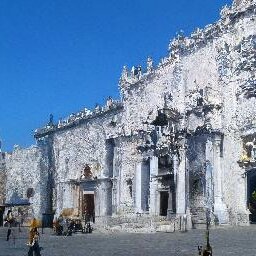}
            \end{tabular}
        \end{minipage}
        \\\noalign{\vskip .1cm}
        \footnotesize \emph{Ours (anisotropic)} (\textbf{FID: 5.7})
        &
        \footnotesize \citet{song2021scorebased} (FID: 14.1)
    \end{tabular}
    \caption{
        Generated samples on \textsc{LSUN\slash church\_outdoor} from our latent space experiment (\cref{sec:experiments-latent-space}). Left: Samples generated by \emph{Ours (anisotropic)} (see \cref{sec:exp_process_aniso_heat_iso_noise}), using a model trained from scratch. Right: Samples generated by \citet{song2021scorebased}, using a model trained from scratch as well. \emph{Ours (anisotropic)} more faithfully resembles the geometric structure of the dataset.
    }
    \label{fig:church-latent}
\end{figure}

These results demonstrate that our anisotropic \gls{spde} framework can be successfully applied in latent space as well. We present generated samples in \cref{fig:church-latent} and \cref{sec:uncurated-samples-lsun-latent,sec:uncurated-samples-bedroom-latent} of the supplemental.

\subsection{Conditional stroke-to-image generation}\label{sec:sdedit}

To demonstrate our model’s ability to better capture geometric structure, we evaluate it in a stroke-guided image generation task using the \textsc{SDEdit} framework~\citep{meng2021sdedit}.
Specifically, we synthetically generate stroke-based paintings from a subset of the dataset via k-means clustering in pixel space. These stroke images are then corrupted using the forward process of each method up to a fixed time horizon, after which the corresponding backward process is applied to generate samples.
We quantitatively compare \emph{Ours (anisotropic)} to \citet{song2021scorebased} by computing the FID between the generated images and the original images from which the stroke paintings were derived. The results show that the introduced anisotropy in \emph{Ours (anisotropic)} consistently outperforms the isotropic baseline of \citet{song2021scorebased} in terms of reconstruction quality.
Quantitative results and representative qualitative examples are shown in \cref{fig:sdedit}.

\section{Conclusion}

Previous diffusion models typically use spatially independent coefficients. We generalize this setting by allowing the coefficients $\anisotropycoefficient_\indexsetelement(\timepoint,\nabla\evolutionpoint)$ to depend on image gradients, thereby introducing state-dependent anisotropy. For $\lambda_\indexsetelement\to\infty$, the coefficients become spatially independent, reducing the model to the classical isotropic case and showing that our framework encompasses isotropic baselines.

We demonstrated that introducing anisotropy into \gls{sbgm} can be practically superior to these traditional isotropic approaches.
%
%
%
%
%
We extended \gls{sbgm} by proposing a novel class of anisotropic diffusion processes theoretically founded on \glspl{spde}. These processes generalize the conventional isotropic setting by enabling geometry-aware transformations that align the generative sampling dynamics more closely with intrinsic geometric structures in the data.

Beyond the theoretical model, we present a proof-of-concept implementation supporting the intuition that the proposed anisotropic framework preserves fine-grained structural information over longer time scales, which is reflected in superior generative performance in both pixel- and latent-space experiments, even with reduced training and inference costs.
Together, these contributions broaden the design space of \glspl{sbgm} and indicate that anisotropic transformations have the potential to further improve sample
quality.

Exploring the broader parameter space of our anisotropic diffusion framework constitutes a promising direction for future work.


\bibliographystyle{ACM-Reference-Format}
\bibliography{bibliography}

\newpage

\clearpage

\begin{table}[H]
    \centering
    \setlength{\tabcolsep}{3pt}
    \small
    \begin{tabular}{lccccccccccccccc}
        \toprule
        & \multicolumn{3}{c}{\textsc{Cifar10}} 
        & \multicolumn{3}{c}{\textsc{CelebA}} 
        & \multicolumn{3}{c}{\textsc{ImageNet2012}} 
        & \multicolumn{3}{c}{\textsc{LSUN/church\_outdoor}} 
        & \multicolumn{3}{c}{\textsc{LSUN/bedroom}} \\
        \cmidrule(lr){2-4} \cmidrule(lr){5-7} \cmidrule(lr){8-10} \cmidrule(lr){11-13} \cmidrule(lr){14-16}
        & IS $\uparrow$ & FID $\downarrow$ & KID $\downarrow$
        & IS $\uparrow$ & FID $\downarrow$ & KID $\downarrow$
        & IS $\uparrow$ & FID $\downarrow$ & KID $\downarrow$
        & IS $\uparrow$ & FID $\downarrow$ & KID $\downarrow$
        & IS $\uparrow$ & FID $\downarrow$ & KID $\downarrow$ \\
        \midrule
        \textbf{Ours (anisotropic)}
        & \textbf{10.2} & \textbf{2.0} & \textbf{6.1e-4}
        & \textbf{3.1} & 2.4 & 1.7e-3
        & \textbf{13.5} & \textbf{19.1} & \textbf{1.9e-2}
        & \textbf{3.6} & \textbf{5.9} & \textbf{4.6e-3}
        & \textbf{3.7} & \textbf{6.1} & \textbf{5.2e-3} \\
        \citet{lipman2022flow}
        & 9.2 & \textbf{2.0} & 7.1e-4
        & 2.4 & \textbf{2.3} & \textbf{1.4e-3}
        & 10.5 & 26.8 & 3.4e-2
        & 3.4 & 26.2 & 5.4e-2
        & - & - & - \\
        \citet{song2021scorebased}
        & 9.8 & 7.1 & 6.6e-4
        & 2.5 & 3.7 & 2.6e-3
        & 12.3 & 24.0 & 2.5e-2
        & 2.5 & 16.7 & 1.2e-2
        & 2.8 & 14.9 & 1.2e-2 \\
        \citet{vandersanden2026edge}
        & 7.1 & 28.7 & 2.2e-2
        & 2.7 & 12.0 & 8.4e-3
        & - & - & -
        & 3.4 & 49.1 & 4.3e-2
        & - & - & - \\
        \bottomrule
    \end{tabular}

    \caption{
        Quantitative comparison across different datasets.
        Higher IS is better, while lower FID and KID are better.
        \emph{Ours (anisotropic)} refers to the anisotropic stochastic heat equation described in \cref{sec:exp_process_aniso_heat_iso_noise}. 
        In the supplemental, we also compare \citet{rissanen2023heat} vs \emph{Ours (isotropic)}~\cref{sec:exp_stoch_heat_iso_noise} on \textsc{Cifar10}.
    }
    \label{tab:metrics}
\end{table}

\begin{table}[H]
    \centering
    \setlength{\tabcolsep}{3pt}
    \small
    \begin{tabular}{lcccccccccccccccc}
        \toprule
        & \multicolumn{4}{c}{\textsc{Cifar10}} 
        & \multicolumn{4}{c}{\textsc{CelebA}} 
        & \multicolumn{4}{c}{\textsc{ImageNet2012}} 
        & \multicolumn{4}{c}{\textsc{LSUN/church\_outdoor}} \\
        \cmidrule(lr){2-5} \cmidrule(lr){6-9} \cmidrule(lr){10-13} \cmidrule(lr){14-17}
        & P $\uparrow$ & R $\uparrow$ & D $\uparrow$ & C $\uparrow$
        & P $\uparrow$ & R $\uparrow$ & D $\uparrow$ & C $\uparrow$
        & P $\uparrow$ & R $\uparrow$ & D $\uparrow$ & C $\uparrow$
        & P $\uparrow$ & R $\uparrow$ & D $\uparrow$ & C $\uparrow$ \\
        \midrule
        \textbf{Ours (anisotropic)}
        & \textbf{0.82} & \textbf{0.79} & \textbf{1.16} & \textbf{0.97}
        & \textbf{0.87} & \textbf{0.74} & 1.25 & \textbf{0.97}
        & \textbf{0.73} & \textbf{0.71} & \textbf{0.79} & \textbf{0.69}
        & \textbf{0.84} & \textbf{0.35} & \textbf{1.18} & \textbf{0.86} \\

        \citet{lipman2022flow}
        & 0.78 & 0.73 & 1.07 & 0.95
        & 0.82 & 0.71 & 1.08 & 0.95
        & 0.60 & 0.66 & 0.58 & 0.54
        & 0.32 & 0.23 & 0.16 & 0.11 \\

        \citet{song2021scorebased}
        & 0.79 & 0.55 & 1.11 & 0.84
        & 0.85 & 0.65 & \textbf{1.26} & 0.95
        & 0.68 & 0.60 & 0.73 & 0.57
        & 0.75 & 0.19 & 0.81 & 0.63 \\

        \citet{vandersanden2026edge}
        & 0.11 & 0.01 & 0.03 & 0.01
        & 0.79 & 0.48 & 0.93 & 0.81
        & - & - & - & -
        & 0.50 & 0.08 & 0.39 & 0.38 \\
        \bottomrule
    \end{tabular}

    \caption{
        Fidelity--diversity trade-off across datasets using precision (P), recall (R), density (D), and coverage (C).
        Higher is better for all metrics.
        Results show that gains in fidelity do not come at the expense of reduced diversity.
    }
    \label{tab:fidelity-diversity}
\end{table}

\begin{figure}[H]
    \centering
    
\newcommand{\imgcell}[2][]{%
  \adjustbox{valign=m}{\includegraphics[#1]{#2}}%
}
\newcommand{\rowcap}[1]{%
  \rotatebox[origin=c]{90}{\footnotesize #1}%
}
\newcommand{\colcap}[1]{\footnotesize #1}

\setlength{\tabcolsep}{2pt} 
\renewcommand{\arraystretch}{1.0}

\adjustbox{max width=\textwidth}{%
\begin{tabular}{@{}ccc@{\hspace{10pt}}ccc@{}}

\multicolumn{3}{c}{\textsc{LSUN\slash church\_outdoor}}
&
\multicolumn{3}{c}{\textsc{LSUN\slash bedroom}}
\\

\imgcell[width=0.25\textwidth]{figures/sdedit/church/stroke_img_0026.jpg}
& \imgcell[width=0.25\textwidth]{figures/sdedit/church/generated_song_0025.jpg}
& \imgcell[width=0.25\textwidth]{figures/sdedit/new/church/1/church1_1.jpeg}

& \imgcell[width=0.25\textwidth]{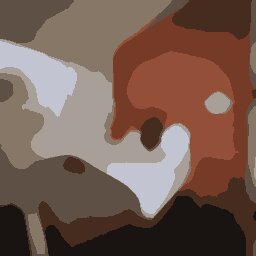}
& \imgcell[width=0.25\textwidth]{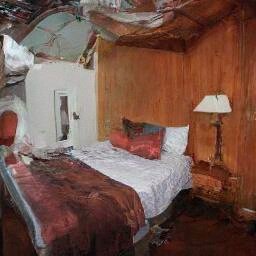}
& \imgcell[width=0.25\textwidth]{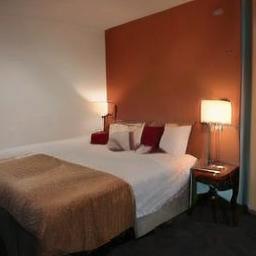}
\\\noalign{\vskip 3pt}

\imgcell[width=0.25\textwidth]{figures/sdedit/church/stroke_img_0356.jpg}
& \imgcell[width=0.25\textwidth]{figures/sdedit/church/generated_song_0355.jpg}
& \imgcell[width=0.25\textwidth]{figures/sdedit/new/church/2/church2_1.jpeg}

& \imgcell[width=0.25\textwidth]{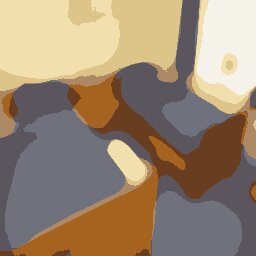}
& \imgcell[width=0.25\textwidth]{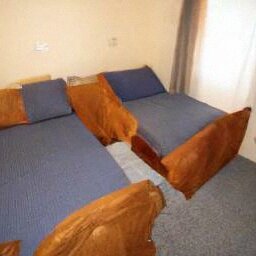}
& \imgcell[width=0.25\textwidth]{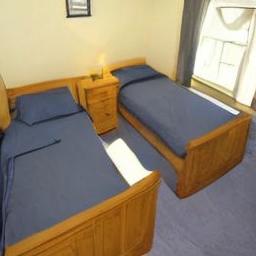}
\\\noalign{\vskip 3pt}

\imgcell[width=0.25\textwidth]{figures/sdedit/church/stroke_img_0521.jpg}
& \imgcell[width=0.25\textwidth]{figures/sdedit/church/generated_song_0520.jpg}
& \imgcell[width=0.25\textwidth]{figures/sdedit/new/church/3/church3_1.jpeg}

& \imgcell[width=0.25\textwidth]{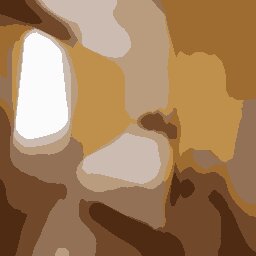}
& \imgcell[width=0.25\textwidth]{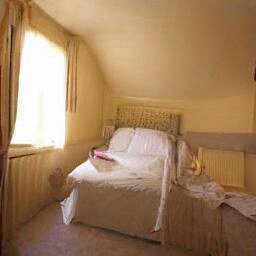}
& \imgcell[width=0.25\textwidth]{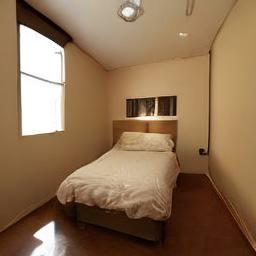}
\\\noalign{\vskip 3pt}

\imgcell[width=0.25\textwidth]{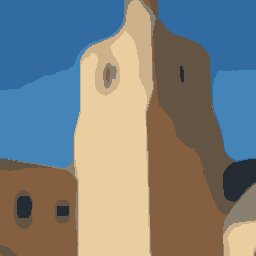}
& \imgcell[width=0.25\textwidth]{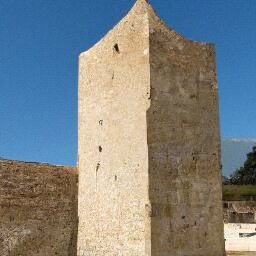}
& \imgcell[width=0.25\textwidth]{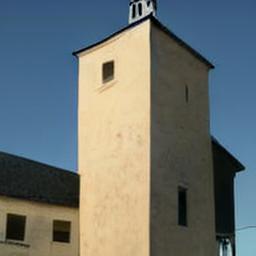}

& \imgcell[width=0.25\textwidth]{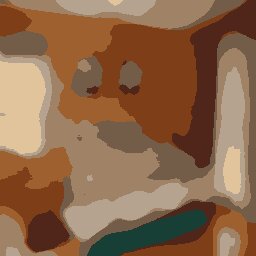}
& \imgcell[width=0.25\textwidth]{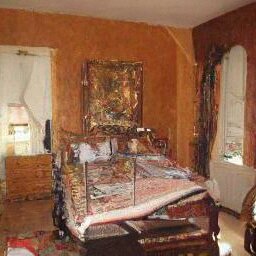}
& \imgcell[width=0.25\textwidth]{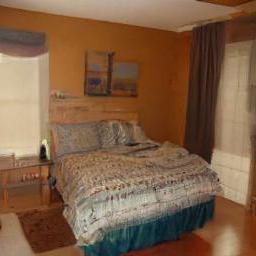}
\\\noalign{\vskip .1cm}

\footnotesize {\def\stackalignment{l}\stackanchor{Stroke painting image}{}}
&
\footnotesize {\def\stackalignment{l}\stackanchor{SDEdit + \citet{song2021scorebased}}{FID: 47.4}}
&
\footnotesize {\def\stackalignment{l}\stackanchor{SDEdit + \emph{Ours (anisotropic)}}{\textbf{FID: 21.9}}}
&
\footnotesize {\def\stackalignment{l}\stackanchor{Stroke painting image}{}}
&
\footnotesize {\def\stackalignment{l}\stackanchor{SDEdit + \citet{song2021scorebased}}{FID: 44.9}}
&
\footnotesize {\def\stackalignment{l}\stackanchor{SDEdit + \emph{Ours (anisotropic)}}{\textbf{FID: 20.2}}}

\\
\end{tabular}%
}

    \caption{
        Stroke-to-image generation with \textsc{SDEdit} (\cref{sec:sdedit}).
        Left block: samples from \textsc{LSUN\slash church\_outdoor}; right block: samples from \textsc{LSUN\slash bedroom}.
        Within each block, columns show (from left to right) the input stroke painting, \textsc{SDEdit} results with \citet{song2021scorebased}, and \textsc{SDEdit} with \emph{Ours (anisotropic)}.}
    \label{fig:sdedit}
\end{figure}

\clearpage

\appendix
\counterwithin{figure}{section}
\counterwithin{table}{section}

\section{Latent space Fidelity--diversity trade-off}\label{sec:latent-space-fd}

In addition to the fidelity--diversity trade-off for the pixel space experiments in \Cref{tab:fidelity-diversity}, we also report precision--recall--density--coverage metrics in \Cref{tab:fidelity-diversity-latent}.

\begin{table*}[t]
    \centering
    \setlength{\tabcolsep}{4pt}
    \small
    \begin{tabular}{lcccccccc}
        \toprule
        & \multicolumn{4}{c}{\textsc{LSUN/church\_outdoor}} 
        & \multicolumn{4}{c}{\textsc{LSUN/bedroom}} \\
        \cmidrule(lr){2-5} \cmidrule(lr){6-9}
        & P $\uparrow$ & R $\uparrow$ & D $\uparrow$ & C $\uparrow$
        & P $\uparrow$ & R $\uparrow$ & D $\uparrow$ & C $\uparrow$ \\
        \midrule
        \textbf{Ours (anisotropic)}
        & \textbf{0.76} & \textbf{0.52} & \textbf{0.89} & \textbf{0.83}
        & \textbf{0.77} & \textbf{0.55} & \textbf{1.31} & \textbf{0.90} \\

        \citet{song2021scorebased}
        & 0.55 & 0.30 & 0.61 & 0.64
        & 0.57 & 0.32 & 0.91 & 0.69 \\
        \bottomrule
    \end{tabular}

    \caption{
        Fidelity--diversity trade-off in latent space using precision (P), recall (R), density (D), and coverage (C).
        Higher is better for all metrics.
        Results remain consistent with pixel-space experiments, indicating that improved fidelity does not come at the expense of reduced diversity.
    }
    \label{tab:fidelity-diversity-latent}
\end{table*}

\section{Design guideline: Controlling image transformations in our framework}\label{sec:design-guide}

In this section, we provide design guidelines by detailing how the individual ingredients of our anisotropic diffusion framework control the image transformations described by our forward process.


\subsection{How drift and diffusion coefficient contribute to degrading information}

The drift $\drift$ \cref{eq:forward-drift} introduces deterministic smoothing that respects the anisotropy defined by the \emph{anisotropy coefficient} $\anisotropytransition_1$. This smoothing is controlled by the \emph{diffusivity coefficient} $\diffusivitytransition_1$, with its strength and direction determined by the local image structure (captured by the gradients $\nabla\forwardprocess_\timepoint$).

The diffusion coefficient $\diffusioncoefficient$ \cref{eq:forward-diffusion-coefficient} injects random noise into the image, which is also modulated anisotropically by $\anisotropytransition_2$. The intensity of the noise is modulated by the \emph{intensity coefficient} $\diffusivitytransition_2$ and destroys fine-grained details in a controlled manner, complementing the drift’s smoothing effect.

While the drift $\drift$ \cref{eq:forward-drift} focuses on smoothing (deterministic destruction), the diffusion coefficient $\diffusioncoefficient$ introduces stochasticity (random destruction).
Together they degrade information in a controlled manner, providing an excellent framework for \gls{sbgm}.

To enable (stochastic) reconstruction, we need to make sure that the forward process destroys the data sufficiently. Anisotropy can be seen as preserving information to some extent. This preserved data can be destroyed in various ways. For example, if the anisotropy is in the drift, we can either destroy the information over time by spreading a lot of isotropic noise at later time points or by transitioning to isotropy in the drift. The latter can be achieved by letting $\anisotropytransition_1(\timepoint)\to\infty$ as $\timepoint\to\timepointmax$.

Subsequently, we will give details on the effect of the diffusivity coefficients $\diffusivitytransition_\sequenceindexmax$ (\cref{sec:diffusivity-coefficient}), intensity coefficient (\cref{sec:intensity-coefficient})
and
anisotropy coefficient $\anisotropytransition_\indexsetelement$ (\cref{sec:anisotropy-coefficient})

\subsubsection{Diffusivity coefficient $\diffusivitytransition_1$}\label{sec:diffusivity-coefficient}


\begin{figure}[H]
    \centering
    \begin{subfigure}[t]{.48\textwidth}
        \centering
        \begin{minipage}[t]{.32\textwidth}
            \setlength{\abovecaptionskip}{2pt}
            \setlength{\belowcaptionskip}{0pt}
            \centering
            \includegraphics[width=\linewidth]{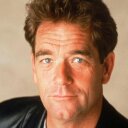}
            \caption*{$\diffusivitytransition_1=0$}
        \end{minipage}
        \hfill
        \begin{minipage}[t]{.32\textwidth}
            \setlength{\abovecaptionskip}{2pt}
            \setlength{\belowcaptionskip}{0pt}
            \centering
            \includegraphics[width=\linewidth]{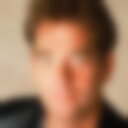}
            \caption*{$\diffusivitytransition_1=1$}
        \end{minipage}
        \hfill
        \begin{minipage}[t]{.32\textwidth}
            \setlength{\abovecaptionskip}{2pt}
            \setlength{\belowcaptionskip}{0pt}
            \centering
            \includegraphics[width=\linewidth]{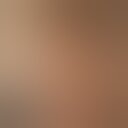}
            \caption*{$\diffusivitytransition_1=.5\to1k$}
        \end{minipage}
        \caption{
            Impact of the diffusivity coefficient $\diffusivitytransition_1$.
            The third column uses a \emph{geometric} transition $\diffusivitytransition_1(\timepoint)=\diffusivitytransition_1^{\textnormal{min}}\left(\diffusivitytransition_1^{\textnormal{max}}/\diffusivitytransition_1^{\textnormal{min}}\right)^{\timepoint/\timepointmax}$ with $\diffusivitytransition_1^{\textnormal{min}}=.5$ and $\diffusivitytransition_1^{\textnormal{max}}=1k$.
            In all columns, we disabled the effect of the diffusion coefficient $\diffusioncoefficient$ by setting $\diffusivitytransition_2=0$.
            %
        }
        \label{fig:diffusion-diffusivity-coefficient-comparison-small}
    \end{subfigure}
    \hfill
    \begin{subfigure}[t]{.48\textwidth}
        \setlength{\abovecaptionskip}{2pt}
            \setlength{\belowcaptionskip}{0pt}
        \centering
        \begin{minipage}[t]{.32\textwidth}
            \centering
            \includegraphics[width=\linewidth]{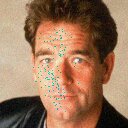}
            \caption*{$\diffusivitytransition_2=1\mathrm{e{-}2}$}
        \end{minipage}
        \hfill
        \begin{minipage}[t]{.32\textwidth}
            \setlength{\abovecaptionskip}{2pt}
            \setlength{\belowcaptionskip}{0pt}
            \centering
            \includegraphics[width=\linewidth]{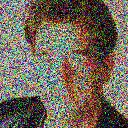}
            \caption*{$\diffusivitytransition_2=1\mathrm{e{-}1}$}
        \end{minipage}
        \hfill
        \begin{minipage}[t]{.32\textwidth}
            \setlength{\abovecaptionskip}{2pt}
            \setlength{\belowcaptionskip}{0pt}
            \centering
            \includegraphics[width=\linewidth]{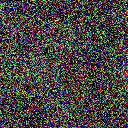}
            \caption*{$\diffusivitytransition_2=1$}
        \end{minipage}
        \captionsetup{skip = 6pt}
        \caption{
            Impact of the intensity coefficient $\diffusivitytransition_2$.
            In all columns, we disabled the effect of the drift $\drift$ by setting $\diffusivitytransition_1=0$.
        }
        \label{fig:diffusion-intensity-coefficient-comparison-small}
    \end{subfigure}
\end{figure}


The diffusivity coefficient $\diffusivitytransition_1$ crucially controls the rate of the (an-)isotropic smoothing in the drift \cref{eq:forward-drift} of the \gls{spde} \cref{eq:forward-equation}.

In the drift \cref{eq:forward-drift}, a larger $\diffusivitytransition_1$ results in stronger smoothing, leading to a faster elimination of high-frequency details (e.g., geometric structure, like edges and corners, and textures) in the image. Conversely, smaller values of $\diffusivitytransition_1$ preserve more of the fine-grained details, slowing down the destruction of information.

In Figure~(\subref{fig:diffusion-diffusivity-coefficient-comparison-small}) we visualized the impact of $\diffusivitytransition_1$, disabling the effect of the diffusion coefficient $\diffusioncoefficient$ by setting $\diffusivitytransition_2=0$. The third column is using a \emph{geometric} transition $\smash{\diffusivitytransition_1(\timepoint)=\diffusivitytransition_1^{\textnormal{min}}\left(\diffusivitytransition_1^{\textnormal{max}}/\diffusivitytransition_1^{\textnormal{min}}\right)^{\timepoint/\timepointmax}}$ with $\diffusivitytransition_1^{\textnormal{min}}=.5$ and $\diffusivitytransition_1^{\textnormal{max}}=1k$. In \hyperref[fig:trans_fns_divergence_fns_vis]{\cref*{fig:trans_fns_divergence_fns_vis} (b)} we show more choices for the diffusivity/intensity coefficients. The original image had a resolution of 128x128 pixels and the forward process was simulated up to time $\timepointmax=8$.

\subsubsection{Intensity coefficient $\diffusivitytransition_2$}\label{sec:intensity-coefficient}

The intensity coefficient $\diffusivitytransition_2$ determines the intensity of the injected noise in the diffusion coefficient \cref{eq:forward-diffusion-coefficient} of the \gls{spde}~\cref{eq:forward-equation}.

In the diffusion coefficient \cref{eq:forward-diffusion-coefficient}, larger $\diffusivitytransition_2$ increases the randomness in the image transformation, introducing more noise and accelerating the destruction of structured information. On the other hand, smaller $\diffusivitytransition_2$ reduces the randomness, preserving some of the original structure while still degrading information.

In \cref{fig:diffusion-intensity-coefficient-comparison-small} we visualized the impact of $\diffusivitytransition_2$, disabling the effect of the drift $\drift$ by setting $\diffusivitytransition_1=0$.

\paragraph*{Impact on generative modeling}

The progression of the $\diffusivitytransition_\indexsetelement$ over time determines how quickly the image is degraded by the forward process.

By modulating $\diffusivitytransition_\indexsetelement$, we can design a destruction process that aligns with the objective of generative modeling: creating a sequence of progressively smoother and noisier images while maintaining enough structure for the model to learn the reconstruction of meaningful samples with a high resemblance to the images from the dataset.

A suitable balance between $\diffusivitytransition_1$ in the drift \cref{eq:forward-drift} and $\diffusivitytransition_2$ in the diffusion coefficient \cref{eq:forward-diffusion-coefficient} ensures that the degradation is smooth but irreversible, providing a structured data destruction trajectory for training the generative model.
\raggedbottom

\subsubsection{Anisotropy coefficient $\anisotropytransition_\indexsetelement$}\label{sec:anisotropy-coefficient}


\begin{figure}[H]
    \vspace{2pt}
    \centering
    \begin{subfigure}[t]{.48\textwidth}
        \centering
        \begin{minipage}[t]{.32\textwidth}
            \setlength{\abovecaptionskip}{2pt}
            \setlength{\belowcaptionskip}{0pt}
            \centering
            \includegraphics[width=\linewidth]{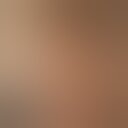}
            \caption*{$\anisotropytransition_1=\infty$}
        \end{minipage}
        \hfill
        \begin{minipage}[t]{.32\textwidth}
            \setlength{\abovecaptionskip}{2pt}
            \setlength{\belowcaptionskip}{0pt}
            \centering
            \includegraphics[width=\linewidth]{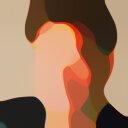}
            \caption*{$\anisotropytransition_1=1\mathrm{e-}2$}
        \end{minipage}
        \hfill
        \begin{minipage}[t]{.32\textwidth}
            \setlength{\abovecaptionskip}{2pt}
            \setlength{\belowcaptionskip}{0pt}
            \centering
            \includegraphics[width=\linewidth]{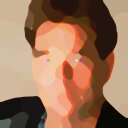}
            \caption*{$\anisotropytransition_1=1\mathrm{e-}3$}
        \end{minipage}
        \caption{
            Impact of the anisotropy coefficient $\diffusivitytransition_1$.    
            In all columns, we disabled the effect of the diffusion coefficient $\diffusioncoefficient$ by setting $\diffusivitytransition_2=0$.
        }
        \label{fig:diffusion-drift-anisotropy-coefficient-small}
    \end{subfigure}
    \hfill
    \begin{subfigure}[t]{.48\textwidth}
        \setlength{\abovecaptionskip}{2pt}
            \setlength{\belowcaptionskip}{0pt}
        \centering
        \begin{minipage}[t]{.32\textwidth}
            \centering
            \includegraphics[width=\linewidth]{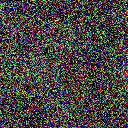}
            \caption*{$\anisotropytransition_2=\infty$}
        \end{minipage}
        \hfill
        \begin{minipage}[t]{.32\textwidth}
            \setlength{\abovecaptionskip}{2pt}
            \setlength{\belowcaptionskip}{0pt}
            \centering
            \includegraphics[width=\linewidth]{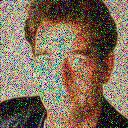}
            \caption*{$\anisotropytransition_2=1\mathrm{e-}2$}
        \end{minipage}
        \hfill
        \begin{minipage}[t]{.32\textwidth}
            \setlength{\abovecaptionskip}{2pt}
            \setlength{\belowcaptionskip}{0pt}
            \centering
            \includegraphics[width=\linewidth]{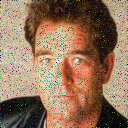}
            \caption*{$\anisotropytransition_2=1\mathrm{e-}3$}
        \end{minipage}
        \caption{
            Impact of the anisotropy coefficient $\diffusivitytransition_2$.
            In all columns, we disabled the effect of the drift $\drift$ by setting $\diffusivitytransition_1=0$.
        }
        \label{fig:diffusion-diffusion-coefficient-anisotropy-coefficient-small}
    \end{subfigure}
\end{figure}


The anisotropy coefficients $\anisotropytransition_\indexsetelement$ control the directional sensitivity of both the drift and diffusion terms by modulating the degree of anisotropy in the transformation. The drift term \cref{eq:forward-drift} depends on $\anisotropycoefficient_1\left(\timepoint,\nabla\forwardprocess_\timepoint\right)$, which introduces directional weighting to the smoothing property of the drift.

When $\anisotropytransition_1$ is small, the smoothing from the drift term \cref{eq:forward-drift} is highly anisotropic, meaning the process prefers certain directions for smoothing while preserving others (e.g., along edges in the image). This ensures that geometric structures are degraded in a structured manner. As $\anisotropytransition_1$ grows larger, the smoothing becomes more isotropic, uniformly degrading all directions and gradually eliminating all structural features.

The diffusion coefficient term \cref{eq:forward-diffusion-coefficient}, modulated by $\anisotropycoefficient_2(\timepoint,\nabla\forwardprocess_\timepoint)$, introduces anisotropic noise. For smaller $\anisotropytransition_2$, the noise is injected
along specific directions, preserving certain patterns while destroying others. As $\anisotropytransition_2$ increases, the noise becomes isotropic, introducing randomness uniformly across the image and further accelerating 
structured information loss.
In the limit, when $\anisotropytransition_2=\infty$, noise is injected isotropically.

In \cref{fig:diffusion-drift-anisotropy-coefficient-small} and \cref{fig:diffusion-diffusion-coefficient-anisotropy-coefficient-small} we visualized the impact of $\anisotropytransition_1$ and $\anisotropytransition_2$, disabling the effect of the diffusion coefficient $\diffusioncoefficient$ and the drift $\drift$ by setting $\diffusivitytransition_2=0$ and $\diffusivitytransition_1=0$, respectively. In \hyperref[fig:trans_fns_divergence_fns_vis]{\cref*{fig:trans_fns_divergence_fns_vis} (a)} we depicted common choices for these anisotropy coefficients.


\paragraph*{Impact on generative modeling}

The anisotropy coefficients $\anisotropytransition_\indexsetelement$ allow for a structured destruction of information. For example, by preserving geometric structures, like edges or corners, for longer time, the forward process provides richer intermediate representations, which can enhance the model’s ability to reconstruct these structures during sampling.

Structured anisotropic degradation may lead to better generative sampling, as the score-based model learns to reverse transformations that align with natural image statistics (e.g., edge preservation and texture destruction).

In contrast, overly isotropic processes (corresponding to large or infinite $\anisotropytransition_\indexsetelement$) degrade the images uniformly, which may simplify the forward process but could result in reduced resemblance of the dataset images, especially if they admit significant geometric patterns.

\section{Incorporating anisotropy in Flow Matching (FM)}\label{sec:anisotropic-flow-matching}

Flow matching methods \citep{lipman2022flow} offer a flexible framework for generative modeling by constructing probability paths between distributions without requiring explicit SPDE formulations. In particular, they support structured perturbations via non-isotropic noise, as demonstrated in edge-aware extensions such as \citep{vandersanden2024edge}. However, these perturbations remain spatially uniform and are typically conditioned only on the initial dataset sample. This limits their ability to adapt to the evolving geometry of the sample during generation.

In contrast, our \gls{spde}-based framework allows for spatially dependent, anisotropic diffusion that reacts dynamically to the evolving image gradients. As a result, structural features such as edges are preserved not only based on their presence in the initial image but also as they emerge, weaken, or shift throughout the transformation. For example, if an initially weak edge becomes stronger over time, our formulation naturally reduces diffusion across it. Conversely, if a previously strong edge fades, the diffusion increases, enabling appropriate smoothing. In flow matching, by contrast, structural information is fixed at the start of the transformation and cannot adapt to changes during generation. This may result in the preservation of features that should fade away or in the undesired blurring of structures that only become salient later in the process.

\section{Conditional generative modeling through anisotropic diffusion bridges}\label{sec:anisotropic-diffusion-brdige}

\gls{ddbm} \citep{zhou2023ddbm} address a fundamentally different task from ours. Rather than generating images unconditionally from noise, \gls{ddbm} learn mappings between two given image distributions by modeling the bridge dynamics connecting them. While their method can, in principle, be applied to unconditional generation by choosing noise as the source distribution, their framework is primarily designed for conditional tasks such as image-to-image translation or editing. In this first application of our framework, however, we focus exclusively on unconditional image generation and therefore do not include a direct comparison.

Extending our framework to enable bridging between arbitrary source and target distributions, akin to \glspl{ddbm}, is part of future work.

\newpage
\section{Sampling from the backward process}\label{sec:backward-sampling}

If $\left(\overline\forwardprocess_\timepoint\right)_{\timepoint\in\timedomain}$ is a Markov process, sampling in the predictor step is done by following its transition dynamics. This is the case, for example, in \gls{ddpm} and in the situation we encounter in our anisotropic diffusion framework from \cref{sec:diffusion-framework} as well.

In \glspl{sdesbm}, we are in the special situation, where the forward process is the solution to an \gls{sde} of the form \begin{equation}\label{eq:forward-sde}
    \dd{\forwardprocess_\timepoint}=\drift\left(\timepoint,\forwardprocess_\timepoint\right)\dd{\timepoint}+\diffusioncoefficient\left(\timepoint,\forwardprocess_\timepoint\right)\dd{\wienerprocess_\timepoint}\eqforall\timepoint\in\timedomain
\end{equation} for some \setword{$\wienercovariance$-Wiener process}{inline:wiener-covariance}. If a suitable regularity condition \citep{Haussmann1986diffusion} is in place, then \begin{equation}\label{eq:reverse-sde}
    \dd{\overline\forwardprocess_\timepoint}=\overline\drift\left(\timepoint,\overline\forwardprocess_\timepoint\right)\dd{\timepoint}+\overline\diffusioncoefficient\left(\timepoint,\overline\forwardprocess_\timepoint\right)\dd{\wienerprocess_\timepoint}\eqforall\timepoint\in\timedomain,
\end{equation} where \begin{align}
    \overline\drift\left(\timepoint,\evolutionpoint\right)&:=\operatorname{tr}{\rm D}_\evolutionpoint\evolutioncovariance\left(\timepointmax-\timepoint,\evolutionpoint\right)+\evolutioncovariance\left(\timepointmax-\timepoint,\evolutionpoint\right)\score\left(\timepointmax-\timepoint,\evolutionpoint\right)-\drift\left(\timepointmax-\timepoint,\evolutionpoint\right);\\
    \overline\diffusioncoefficient\left(\timepoint,\evolutionpoint\right)&:=\diffusioncoefficient\left(\timepointmax-\timepoint,\evolutionpoint\right)
\end{align} for $(\timepoint,\point)\in\timedomain\times\mathbb R^\discretedomain$ and \begin{equation}
    \evolutioncovariance:=\left(\diffusioncoefficient\wienercovariance^{\frac12}\right)\left(\diffusioncoefficient\wienercovariance^{\frac12}\right)^\ast.
\end{equation} In this case, the sampling in the predictor step is performed using a method for the numerical solution of a \gls{sde}, with the Euler-Maruyama method \citep{kloeden1992numerical} being the simplest approach.

\paragraph*{Discussion}

We emphasize that, in the practical application of our framework, it is the finite-dimensional \gls{sde} — obtained via the numerical scheme simulating our \gls{spde} \cref{eq:forward-equation} described in \cref{sec:numerical-simulation} — that must be reversed in time, not the \gls{spde} \cref{eq:forward-equation} itself. While --- under a suitable set of assumptions --- time-reversal of the \gls{spde} \cref{eq:forward-equation} is theoretically possible \citep{follmer1986timereveral, millet1989reversal}, the results presented in \citet{Haussmann1986diffusion, anderson1982reversal} are sufficient for our purposes, as they apply directly to the finite-dimensional \gls{sde} setting.


\section{
Diffusivity, intensity and anisotropy coefficients $\diffusivitytransition_1$, $\diffusivitytransition_2$ and $\anisotropytransition_\indexsetelement$}

\begin{figure}[H]
    \centering
    \input{figures/transition_and_divergence_fn_vis} 
    
    \caption{(a) Visualization of the common choice $\anisotropytransition(\timepoint):=\anisotropytransition^{\textnormal{min}}\frac{e^{\exponentialparam\timepointmax}-1}{e^{\exponentialparam\left(\timepointmax-\timepoint\right)}-1}$ for the anisotropy coefficients (\cref{sec:anisotropy-coefficient}). (b) Visualization of the common choice $\diffusivitytransition(\timepoint):=\diffusivitytransition^{\textnormal{min}}+\left(\diffusivitytransition^{\textnormal{max}}-\diffusivitytransition^{\textnormal{min}}\right)\left(\frac\timepoint\timepointmax\right)^\powerorder$ for the diffusivity (\cref{sec:diffusivity-coefficient}) and intensity coefficients (\cref{sec:intensity-coefficient}).}
    \label{fig:trans_fns_divergence_fns_vis}
\end{figure}

\section{The Cameron-Martin space $\wienercovariance^{\frac12}L^2(\domain)$}\label{sec:cameron-martin}

The space $\wienercovariance^{\frac12}L^2(\domain)$ is usually called a \emph{Cameron-Martin space}; see \citep[Chapter~I.4]{daprato2014evolution} or \citep[Definition~10.15]{lord2014spde}. To understand and apply our paper, it is only important to be aware of the set-theoretic definition $\wienercovariance^{\frac12}L^2(\domain):=\{\wienercovariance^{\frac12}\evolutionpoint:\evolutionpoint\in L^2(\domain)\}$. In plain English, it is the space of all transformations of $L^2(\domain)$-function under the operator $\wienercovariance^{\frac12}$. For more theoretical considerations, one important aspect is that it inherits a Hilbert space structure from $L^2(\domain)$.

\section{\gls{spde} classification}\label{sec:spde-classification}

In general, \cref{eq:forward-equation} is a quasilinear parabolic \gls{spde} with multiplicative noise.

If $\anisotropycoefficient_1$ does not depend on the second argument, \cref{eq:forward-equation} is a semilinear parabolic \gls{spde} with multiplicative noise: \begin{equation}\label{eq:stochastic-heat-with-multiplicative-noise}
    \dd{\forwardprocess_\timepoint}=\diffusivitytransition_1(\timepoint)\Delta\forwardprocess_\timepoint\dd{\timepoint}+\diffusioncoefficient(\timepoint,\forwardprocess_\timepoint)\dd{\wienerprocess_\timepoint}\eqforall\timepoint\in\timedomain.
\end{equation}

If $\anisotropycoefficient_2$ does not depend on the second argument,  \cref{eq:forward-equation} is a quasilinear parabolic \gls{spde} with additive noise: \begin{equation}\label{eq:anisotropic-stochastic-heat-with-additive-noise}
    \dd{\forwardprocess_\timepoint}=\drift(\timepoint,\forwardprocess_\timepoint)\dd{\timepoint}+\diffusivitytransition_2(\timepoint)\dd{\wienerprocess_\timepoint}\eqforall\timepoint\in\timedomain.
\end{equation}

Finally, if $\anisotropycoefficient_1$ and $\anisotropycoefficient_2$ both do not depend on the second argument, \cref{eq:forward-equation} is a linear parabolic \gls{spde} with additive noise: \begin{equation}\label{eq:stochastic-heat-with-additive-noise}
    \dd{\forwardprocess_\timepoint}=\diffusivitytransition_1(\timepoint)\Delta\forwardprocess_\timepoint\dd{\timepoint}+\diffusivitytransition_2(\timepoint)\dd{\wienerprocess_\timepoint}\eqforall\timepoint\in\timedomain.
\end{equation}


\section{Numerical simulation}\label{sec:numerical-simulation}

For the numerical simulation of the forward and backward processes, \cref{eq:forward-equation} and \cref{eq:backward-process}, we modeled the image space $\domain$ as $\domain=(0,\discretedimension_1)\times(0,\discretedimension_2)$ and decomposed the boundary $\partial\domain$ according to \begin{align}
    \partial_L\domain&:=\{0\}\times[0,\discretedimension_2);\\
    \partial_T\domain&:=[0,\discretedimension_1)\times\{\discretedimension_2\};\\
    \partial_R\domain&:=\{\discretedimension_1\}\times(0,\discretedimension_2];\\
    \partial_B\domain&:=(0,\discretedimension_1]\times\{0\}
\end{align} into its left, top, right and bottom part. We discretized the derivatives using a mixture of forward, backward and central finite differences, respecting Neumann boundary conditions.

\subsection{Domain discretization}\label{sec:domain-discretization}

After discretization, we decomposed the discretized domain $\discretedomain=\left\{0,\ldots,\discretedimension_1\right\}\times\left\{0,\ldots,\discretedimension_2\right\}$ in the same spirit into its interior, left, top, right and bottom part: \begin{align}
    \discretedomain^\circ&:=\{1,\ldots,\discretedimension_1-2\}\times\{1,\ldots,\discretedimension_2-2\};\\
    \partial_L\discretedomain&:=\{0\}\times\{0,\ldots,\discretedimension_2 - 2);\\
    \partial_T\discretedomain&:=\{0,\ldots,\discretedimension_2-2\}\times\{\discretedimension_2-1\};\\
    \partial_R\discretedomain&:=\{\discretedimension_1-1\}\times\{1,\ldots,\discretedimension_2-1\};\\
    \partial_B\discretedomain&:=\{1,\ldots,\discretedimension_1-1\}\times\{0\}.
\end{align}

\clearpage

\subsection{Spatial discretization}

For the finite-difference approximation we have chosen, the discretized drift is given by \begin{equation}
    \tilde\drift\left(\timepoint,\evolutionpoint\right)_\indexsetelement:=\begin{cases}
        \begin{aligned}[b]
            &\anisotropycoefficient_1\left(\timepoint,\begin{pmatrix}\evolutionpoint_{\subscriptindex{\subscriptmin{\indexsetelement_1+2}{\dimension_1-1}}{\indexsetelement_2}}-\dcc\evolutionpoint\indexsetelement\\\dff\evolutionpoint\indexsetelement-\dfb\evolutionpoint\indexsetelement\end{pmatrix}\right)\left(\dfc\evolutionpoint\indexsetelement-\dcc\evolutionpoint\indexsetelement\right)\;\\
            &\quad\quad-\anisotropycoefficient_1\left(\timepoint,\begin{pmatrix}\dcc\evolutionpoint\indexsetelement-\evolutionpoint_{\subscriptindex{\subscriptmax{\indexsetelement_1-2}0}{\indexsetelement_2}}\\\dbf\evolutionpoint\indexsetelement-\dbb\evolutionpoint\indexsetelement\end{pmatrix}\right)\left(\dcc\evolutionpoint\indexsetelement-\dbc\evolutionpoint\indexsetelement\right)\;\\
            &\quad\quad+\anisotropycoefficient_1\left(\timepoint,\begin{pmatrix}\dff\evolutionpoint\indexsetelement-\dbf\evolutionpoint\indexsetelement\\\evolutionpoint_{\subscriptindex{\indexsetelement_1}{\subscriptmin{\indexsetelement_2+2}{\dimension_2-1}}}-\dcc\evolutionpoint\indexsetelement\end{pmatrix}\right)\left(\dcf\evolutionpoint\indexsetelement-\dcc\evolutionpoint\indexsetelement\right)\;\\
            &\quad\quad-\anisotropycoefficient_1\left(\timepoint,\begin{pmatrix}\dfb\evolutionpoint\indexsetelement-\dbb\evolutionpoint\indexsetelement\\\dcc\evolutionpoint\indexsetelement-\evolutionpoint_{\subscriptindex{\indexsetelement_1}{\subscriptmax{\indexsetelement_2-2}0}}\end{pmatrix}\right)\left(\dcc\evolutionpoint\indexsetelement-\dcb\evolutionpoint\indexsetelement\right)
        \end{aligned}&\text{, if }\indexsetelement\in\discretedomain^\circ;\\
        \begin{aligned}[b]
            &\left(\anisotropycoefficient_1\left(\timepoint,\begin{pmatrix}\evolutionpoint_{\subscriptindex{\subscriptmin{\indexsetelement_1+2}{\discretedimension_1-1}}{\indexsetelement_2}}-\evolutionpoint_\indexsetelement\\0\end{pmatrix}\right)+\anisotropycoefficient_1\left(\timepoint,0\right)\right)\left(\dfc\evolutionpoint\indexsetelement-\evolutionpoint_\indexsetelement\right)\\
            &\quad\quad+\left(\anisotropycoefficient_1\left(\timepoint\begin{pmatrix}0\\\evolutionpoint_{\subscriptindex{\indexsetelement_1}{\subscriptmin{\indexsetelement_2+2}{\discretedimension_2-1}}}-\evolutionpoint_\indexsetelement\end{pmatrix}\right)+\anisotropycoefficient_1\left(\timepoint,0\right)\right)\left(\dcf\evolutionpoint\indexsetelement-\evolutionpoint_\indexsetelement\right)
        \end{aligned}&\text{, if }\indexsetelement\in\partial_L\discretedomain\text{ with }\indexsetelement_2=0;\\
        \begin{aligned}[b]
            &\left(\anisotropycoefficient_1\left(\timepoint,\begin{pmatrix}\evolutionpoint_{\subscriptindex{\subscriptmin{\indexsetelement_1+2}{\dimension_1-1}}{\indexsetelement_2}}-\dcc\evolutionpoint\indexsetelement\\\dff\evolutionpoint\indexsetelement-\dfb\evolutionpoint\indexsetelement\end{pmatrix}\right)\right.\\
            &\quad\quad+\left.\anisotropycoefficient_1\left(\timepoint,\begin{pmatrix}0\\\dff\evolutionpoint\indexsetelement-\dfb\evolutionpoint\indexsetelement\end{pmatrix}\right)\right)\left(\dfc\evolutionpoint\indexsetelement-\dcc\evolutionpoint\indexsetelement\right)\\
            &\quad\quad+\anisotropycoefficient_1\left(\timepoint,\begin{pmatrix}0\\\evolutionpoint_{\subscriptindex{\indexsetelement_1}{\subscriptmin{\indexsetelement_2+2}{\discretedomain_2-1}}}-\evolutionpoint_\indexsetelement\end{pmatrix}\right)\left(\dcf\evolutionpoint\indexsetelement-\evolutionpoint_\indexsetelement\right)\\
            &\quad\quad-\anisotropycoefficient_1\left(\timepoint,\begin{pmatrix}0\\\evolutionpoint_\indexsetelement-\evolutionpoint_{\subscriptindex{\indexsetelement_1}{\subscriptmax{\indexsetelement_2-2}0}}\end{pmatrix}\right)\left(\evolutionpoint_\indexsetelement-\dcb\evolutionpoint\index
            \right)
        \end{aligned}&\text{, if }\indexsetelement\in\partial_L\discretedomain\text{ with }\indexsetelement_2>0;\\
        \begin{aligned}[b]
            &\left(\anisotropycoefficient_1\left(\timepoint,\begin{pmatrix}\evolutionpoint_{\subscriptindex{\subscriptmin{\indexsetelement_1+2}{\discretedimension_1-1}}{\indexsetelement_2}}-\evolutionpoint_\indexsetelement\\0\end{pmatrix}\right)+\anisotropycoefficient_1\left(\timepoint,0\right)\right)\left(\dfc\evolutionpoint\indexsetelement-\evolutionpoint_\indexsetelement\right)\\
            &\quad\quad+\left(\anisotropycoefficient_1\left(\timepoint,\begin{pmatrix}0\\\evolutionpoint_\indexsetelement-\evolutionpoint_{\subscriptindex{\indexsetelement_1}{\subscriptmax{\indexsetelement_2}0}}\end{pmatrix}\right)+\anisotropycoefficient_1\left(\timepoint,0\right)\right)\left(\dcb\evolutionpoint\indexsetelement-\evolutionpoint_\indexsetelement\right)
        \end{aligned}&\text{, if }\indexsetelement\in\partial_T\discretedomain\text{ with }\indexsetelement_1=0;\\
        \begin{aligned}[b]
            &\anisotropycoefficient_1\left(\timepoint,\begin{pmatrix}\evolutionpoint_{\subscriptindex{\subscriptmin{\indexsetelement_1+2}{\discretedimension_1-1}}{\indexsetelement_2}}-\evolutionpoint_\indexsetelement\\0\end{pmatrix}\right)\\
            &\quad\quad-\anisotropycoefficient_1\left(\timepoint,\begin{pmatrix}\evolutionpoint_\indexsetelement-\evolutionpoint_{\subscriptindex{\subscriptmax{\indexsetelement_1-2}0}{\indexsetelement_2}}\\0\end{pmatrix}\right)\left(\evolutionpoint_\indexsetelement-\dbc\evolutionpoint\indexsetelement\right)\\
            &\quad\quad+\left(\anisotropycoefficient_1\left(\timepoint,\begin{pmatrix}\dfb\evolutionpoint\indexsetelement-\dbb\evolutionpoint\indexsetelement\\0\end{pmatrix}\right)\right.\\
            &\quad\quad+\left.\anisotropycoefficient_1\left(\begin{pmatrix}\dfb\evolutionpoint\indexsetelement-\dbb\evolutionpoint\indexsetelement\\\evolutionpoint_\indexsetelement-\evolutionpoint_{\subscriptindex{\indexsetelement_1}{\subscriptmax{\indexsetelement_2-2}0}}\end{pmatrix}\right)\right)\left(\dcb\evolutionpoint\indexsetelement-\evolutionpoint_\indexsetelement\right)
        \end{aligned}&\text{, if }\indexsetelement\in\partial_T\discretedomain\text{ with }\indexsetelement_1>0
    \end{cases}
\end{equation} and \begin{equation}
    \tilde\drift\left(\timepoint,\evolutionpoint\right)_\indexsetelement:=\begin{cases}
        \begin{aligned}[b]
            &\left(\anisotropycoefficient_1\left(\timepoint,\begin{pmatrix}\evolutionpoint_\indexsetelement-\evolutionpoint_{\subscriptindex{\subscriptmax{\indexsetelement_1-2}0}{\indexsetelement_2}}\\0\end{pmatrix}\right)+\anisotropycoefficient_1\left(\timepoint,0\right)\right)\left(\dbf\evolutionpoint\indexsetelement-\evolutionpoint_\indexsetelement\right)\\
            &\quad\quad+\left(\anisotropycoefficient_1\left(\timepoint,\begin{pmatrix}0\\\evolutionpoint_\indexsetelement-\evolutionpoint_{\subscriptindex{\indexsetelement_1}{\subscriptmax{\indexsetelement_2-2}0}}\end{pmatrix}\right)+\anisotropycoefficient_1\left(\timepoint,0\right)\right)\left(\dcb\evolutionpoint\indexsetelement-\evolutionpoint_\indexsetelement\right)
        \end{aligned}&\text{, if }\indexsetelement\in\partial_R\discretedomain\text{ with }\indexsetelement_2=\discretedimension_2-1;\\
        \begin{aligned}[b]
            &\left(\anisotropycoefficient_1\left(\timepoint,\begin{pmatrix}0\\\dbf\evolutionpoint\indexsetelement-\dbb\evolutionpoint\indexsetelement\end{pmatrix}\right)\right.\\
            &\quad\quad+\left.\anisotropycoefficient_1\left(\timepoint,\begin{pmatrix}\evolutionpoint_\indexsetelement-\evolutionpoint_{\subscriptindex{\subscriptmax{\indexsetelement_1-2}0}{\indexsetelement_2}}\\\dbf\evolutionpoint\indexsetelement-\dbb\evolutionpoint\indexsetelement\end{pmatrix}\right)\right)\left(\dbc\evolutionpoint\indexsetelement-\evolutionpoint_\indexsetelement\right)\\
            &\quad\quad+\anisotropycoefficient_1\left(\timepoint,\begin{pmatrix}0\\\evolutionpoint_{\subscriptindex{\indexsetelement_1}{\subscriptmin{\indexsetelement_2+2}{\discretedimension_2-1}}}-\evolutionpoint_\indexsetelement\end{pmatrix}\right)\left(\dcf\evolutionpoint\indexsetelement-\evolutionpoint_\indexsetelement\right)\\
            &\quad\quad-\anisotropycoefficient_1\left(\timepoint,\begin{pmatrix}0\\\evolutionpoint_\indexsetelement-\evolutionpoint_{\subscriptindex{\indexsetelement_1}{\subscriptmax{\indexsetelement_2-2}0}}\end{pmatrix}\right)\left(\evolutionpoint_\indexsetelement-\dcb\evolutionpoint\indexsetelement\right)
        \end{aligned}&\text{, if }\indexsetelement\in\partial_R\discretedomain\text{ with }\indexsetelement_2<\discretedimension_2-1;\\
        \begin{aligned}[b]
            &\left(\anisotropycoefficient_1\left(\timepoint,\begin{pmatrix}\evolutionpoint_\indexsetelement-\evolutionpoint_{\subscriptindex{\subscriptmax{\indexsetelement_1-2}0}{\indexsetelement_2}}\\0\end{pmatrix}\right)+\anisotropycoefficient_1\left(\timepoint,0\right)\right)\left(\dbc\evolutionpoint\indexsetelement-\evolutionpoint_\indexsetelement\right)\\
            &\quad\quad+\left(\anisotropycoefficient_1\left(\timepoint,\begin{pmatrix}0\\\evolutionpoint_{\subscriptindex{\indexsetelement_1}{\subscriptmin{\indexsetelement_2+2}{\discretedimension_2-1}}}-\evolutionpoint_\indexsetelement\end{pmatrix}\right)+\anisotropycoefficient_1\left(\timepoint,0\right)\right)\left(\dcf\evolutionpoint\indexsetelement-\evolutionpoint_\indexsetelement\right)
        \end{aligned}&\text{, if }\indexsetelement\in\partial_B\discretedomain\text{ with }\indexsetelement_1=\discretedimension_1-1;\\
        \begin{aligned}[b]
            &\anisotropycoefficient_1\left(\timepoint,\begin{pmatrix}\evolutionpoint_{\subscriptindex{\subscriptmin{\indexsetelement_1+2}{\discretedimension_1-1}}{\indexsetelement_2}}-\evolutionpoint_\indexsetelement\\0\end{pmatrix}\right)\left(\dfc\evolutionpoint\indexsetelement-\evolutionpoint_\indexsetelement\right)\\
            &\quad\quad-\anisotropycoefficient_1\left(\timepoint,\begin{pmatrix}\evolutionpoint_\indexsetelement-\evolutionpoint_{\subscriptindex{\subscriptmax{\indexsetelement_1-2}0}{\indexsetelement_2}}\\0\end{pmatrix}\right)\left(\evolutionpoint_\indexsetelement-\dbc\evolutionpoint\indexsetelement\right)\\
            &\quad\quad+\left(\anisotropycoefficient_1\left(\timepoint,\begin{pmatrix}\dff\evolutionpoint\indexsetelement-\dbf\evolutionpoint\indexsetelement\\\evolutionpoint_{\subscriptindex{\indexsetelement_1}{\subscriptmin{\indexsetelement_2+2}{\discretedimension_2-1}}}-\evolutionpoint_\indexsetelement\end{pmatrix}\right)\right.\\
            &\quad\quad+\left.\anisotropycoefficient_1\left(\timepoint,\begin{pmatrix}\dff\evolutionpoint\indexsetelement-\dbf\evolutionpoint\indexsetelement\\0\end{pmatrix}\right)\right)\left(\dcf\evolutionpoint\indexsetelement-\evolutionpoint_\indexsetelement\right)
        \end{aligned}&\text{, if }\indexsetelement\in\partial_B\discretedomain\text{ with }\indexsetelement_1<\discretedimension_1-1
    \end{cases}
\end{equation} for $(\timepoint,\evolutionpoint)\in\timedomain\times\mathbb R^\discretedomain$ and $\indexsetelement\in\discretedomain$ and the discretized diffusion coefficient being given by \begin{equation}\label{eq:discretized-diffusion-coefficient}
    \left(\tilde\diffusioncoefficient\left(\timepoint,\evolutionpoint\right)\noisemode\right)_\indexsetelement:=\left.\begin{cases}
    \anisotropycoefficient_2\left(\timepoint,\begin{pmatrix}\evolutionpoint_{\subscriptindex{\indexsetelement_1+1}{\indexsetelement_2}}-\evolutionpoint_{\indexsetelement_1-1,\indexsetelement_2}\\\evolutionpoint_{\indexsetelement_1,\indexsetelement_2+1}-\evolutionpoint_{\indexsetelement_1,\indexsetelement_2-1}\end{pmatrix}\right)&\text{, if }\indexsetelement\in\discretedomain^\circ;\\
        \anisotropycoefficient_2\left(t,\begin{pmatrix}0\\\evolutionpoint_{\indexsetelement_1,\indexsetelement_2+1}-\evolutionpoint_{\indexsetelement_1,\indexsetelement_2-1}\end{pmatrix}\right)&\text{, if }\begin{aligned}[t]&\indexsetelement\in\partial_L\discretedomain\text{ with }\indexsetelement_2>0\text{ or }\\&\indexsetelement\in\partial_R\discretedomain\text{ with }\indexsetelement_2<\discretedimension_2-1;\end{aligned}\\
        \anisotropycoefficient_2\left(t,\begin{pmatrix}\evolutionpoint_{\indexsetelement_1+1,\indexsetelement_2}-\evolutionpoint_{\indexsetelement_1-1,\indexsetelement_2}\\0\end{pmatrix}\right)&\text{, if }\begin{aligned}[t]&\indexsetelement\in\partial_T\discretedomain\text{ with }\indexsetelement_1>0\text{ or }\\&\indexsetelement\in\partial_B\discretedomain\text{ with }\indexsetelement_1<\discretedimension_1-1;\end{aligned}\\
        \anisotropycoefficient_2(t,0)&\text{, if }\begin{aligned}[t]&\indexsetelement\in\partial_L\discretedomain\text{ with }\indexsetelement_2=0\text{ or }\\&\indexsetelement\in\partial_T\discretedomain\text{ with }\indexsetelement_1=0\text{ or }\\&\indexsetelement\in\partial_R\discretedomain\text{ with }\indexsetelement_2=\discretedimension_2-1\text{ or }\\&\indexsetelement\in\partial_B\discretedomain\text{ with }\indexsetelement_1=\discretedimension_1-1\end{aligned}
    \end{cases}\right\}\noisemode_\indexsetelement
\end{equation} for $\noisemode\in\mathbb R^\discretedomain$, $(\timepoint,\evolutionpoint)\in\timedomain\times\mathbb R^\discretedomain$ and $\indexsetelement\in\discretedomain$.

\subsection{Temporal discretization}

For temporal discretization, we used a \emph{drift-termed} (explicit) Euler-Maruyama scheme \citep{hutzenthaler2015numerical}, where given a generic \gls{sde} of the form \begin{equation}\label{eq:generic-sde}
    \dd{\tilde\forwardprocess_\timepoint}=\tilde\drift\left(\timepoint,\tilde\forwardprocess_\timepoint\right)\dd{\timepoint}+\tilde\diffusioncoefficient\left(\timepoint,\tilde\forwardprocess_\timepoint\right)\dd{\tilde\wienerprocess_\timepoint}\eqforall\timepoint\in\timedomain,
\end{equation} the time stepping is given by \begin{equation}
    \tilde\forwardprocess_{\timepoint+\Delta\timepoint}=\tilde\forwardprocess_\timepoint+\frac{\drift\left(\timepoint,\tilde\forwardprocess_\timepoint\right)}{1+\Delta\timepoint\left\|\drift\left(\timepoint,\tilde\forwardprocess_\timepoint\right)\right\|^\tamingcoefficient}+\tilde\diffusioncoefficient\left(\timepoint,\tilde\forwardprocess_\timepoint\right)\left(\tilde\wienerprocess_{\timepoint+\Delta\timepoint}-\tilde\wienerprocess_\timepoint\right)
\end{equation} for all $\timepoint,\Delta\timepoint\ge0$ with $\timepoint+\Delta\timepoint\in\timedomain$, where $\gamma$ is a \emph{taming coefficient} usually chosen to be $1$.

\begin{table}[b]
    \centering
    \setlength{\tabcolsep}{5pt}
    \small
    \begin{tabular}{lccc}
        \toprule
        & IS $\uparrow$ & FID $\downarrow$ & KID $\downarrow$ \\
        \midrule
        \textbf{Ours (isotropic)} & \textbf{8.8} & \textbf{19.6} & \textbf{1.6e-02} \\
        \citet{rissanen2023heat} & 5.9 & 84.3 & 7.2e-2 \\
        \bottomrule
    \end{tabular}
    \caption{
        Comparison on \textsc{Cifar10}. \emph{Ours (isotropic)} achieves significantly better FID and KID compared to \citet{rissanen2023heat}. 
        \emph{Ours (isotropic)} refer to the isotropic  stochastic heat equation with isotropic noise described in Sec.~5.3 from the main paper.
    }
    \label{tab:cifar_iso}
\end{table}

\section{\gls{spde} trajectory visualizations}

\begin{figure}[H]
    \centering
    \input{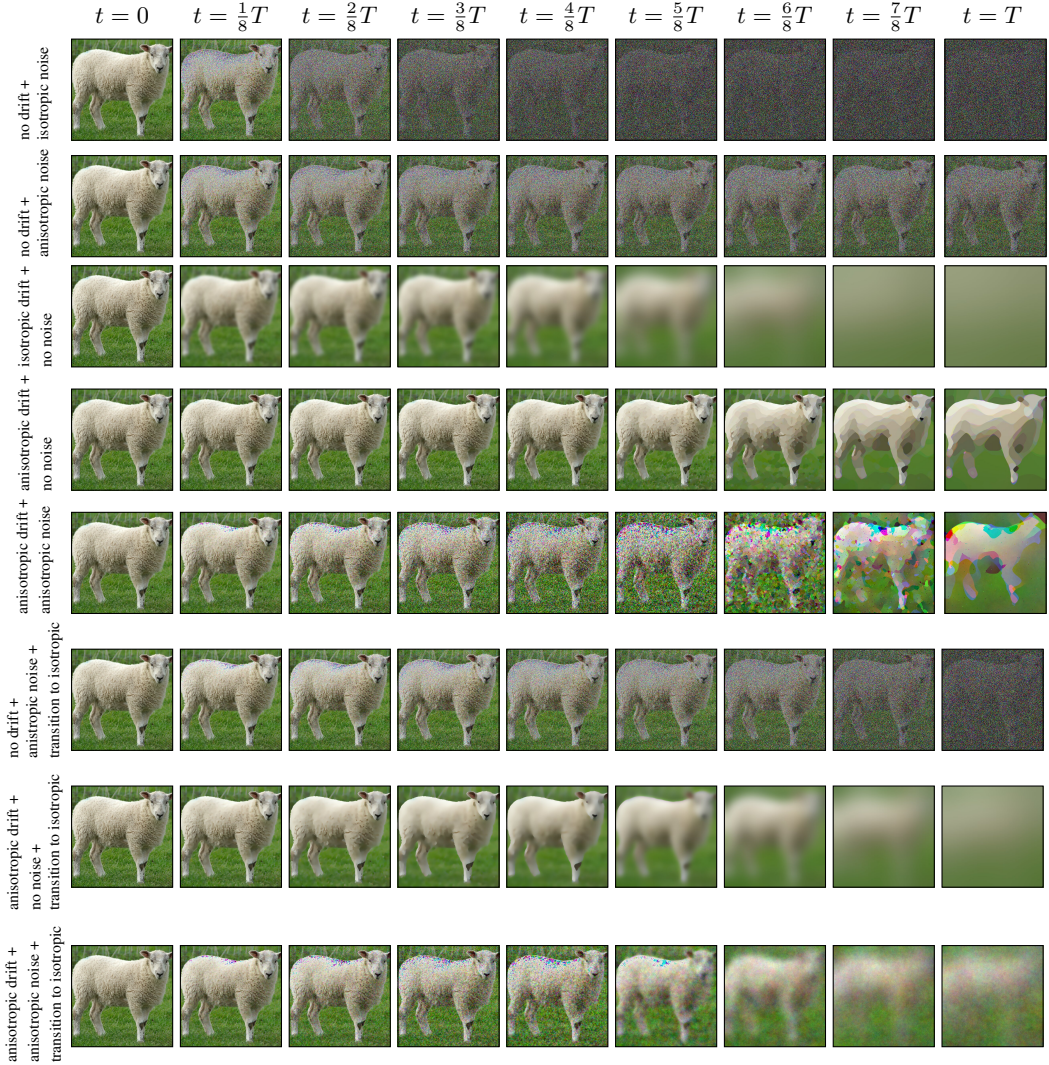} 
    \caption{We visualize the core ingredients of our generalized anisotropic \gls{spde} diffusion framework. The diffusion process is governed by two fundamental components: the drift term, driven by the drift coefficient $\drift$ and the diffusion term, driven by the diffusion coefficient $\diffusioncoefficient$. Both terms can take on isotropic or anisotropic forms, and their combinations open the door to a vast spectrum of processes. These processes destroy (and regenerate, for the reverse process) the signal’s information in ways that range from subtle to profoundly distinct. The interplay between these terms offers the designer of the generative process fine control over how information is destroyed.}
\label{fig:fwd_process_vis}
\end{figure}

\section{Uncurated generated samples on \textsc{Cifar10}}\label{sec:uncurated-samples-cifar10}

\begin{figure}[H]
    \centering

\newcommand{\PlotSingleImage}[1]{%
        \begin{scope}
            \clip (0,0) -- (2.5,0) -- (2.5,2.5) -- (0,2.5) -- cycle;
            \path[fill overzoom image=figures/#1] (0,0) rectangle (2.5cm,2.5cm);
        \end{scope}
        \draw (0,0) -- (2.5,0) -- (2.5,2.5) -- (0,2.5) -- cycle;
        
}

\newcommand{\PlotSingleImageNonSquared}[1]{%
        \begin{scope}
            \clip (0,0) -- (4,0) -- (4,10) -- (0,10) -- cycle;
            \path[fill overzoom image=figures/#1] (0,0) rectangle (4cm,10cm);
        \end{scope}
        \draw (0,0) -- (4,0) -- (4,10) -- (0,10) -- cycle;
        
}

\newcommand\scalevalue{0.80}    

%
\hspace*{-4.0mm}
\begin{tabular}{c@{\;}c@{\;}c@{\;}c@{}}
\begin{tabular}{c@{\;}c@{}}
\begin{tikzpicture}[scale=\scalevalue]
\PlotSingleImageNonSquared{uncurated_samples/uncurated_ihdm.jpg}
\end{tikzpicture}
\\[-0.4mm]
{\def\stackalignment{l}\stackanchor{\citet{rissanen2023heat}}{(FID: 84.3)}}
\\[-0.4mm]
\end{tabular}
&
\begin{tabular}{c@{\;}c@{}}
\begin{tikzpicture}[scale=\scalevalue]
\PlotSingleImageNonSquared{uncurated_samples/uncurated_score_sde.jpg}
\end{tikzpicture}
\\[-0.4mm]

{\def\stackalignment{l}\stackanchor{\citet{song2021scorebased}}{(FID: 7.1)}}
\\[-0.4mm]
\end{tabular}
&
\begin{tabular}{c@{\;}c@{}}
\begin{tikzpicture}[scale=\scalevalue]
\PlotSingleImageNonSquared{uncurated_samples/uncurated_ours_heat.jpg}
\end{tikzpicture}
\\[-0.4mm]

{\def\stackalignment{l}\stackanchor{Ours (isotropic)}{(FID: 19.6)}}
\\[-0.4mm]
\end{tabular}
\begin{tabular}{c@{\;}c@{}}
\begin{tikzpicture}[scale=\scalevalue]
\PlotSingleImageNonSquared{uncurated_samples/uncurated_ours_aniso_v1.jpg}
\end{tikzpicture}
\\[-0.4mm]

{\def\stackalignment{l}\stackanchor{Ours (anisotropic)}{(FID: 2.0)}}
\\[-0.4mm]
\end{tabular}
\end{tabular}
 
    \vspace{-0.5mm}
  
    \caption{Uncurated samples for the baselines \citet{rissanen2023heat}, \citet{song2021scorebased} and two of our \glspl{spde}: one with isotropic drift $\drift$ and isotropic diffusion coefficient $\diffusioncoefficient$ (see \cref{sec:exp_stoch_heat_iso_noise}) and one with anisotropic $\drift$ and isotropic $\diffusioncoefficient$ (see \cref{sec:exp_process_aniso_heat_iso_noise}). Note that \citet{rissanen2023heat} and \emph{Ours (isotropic)} essentially represent the same \gls{spde}, but our score-based approach performs significantly better.}
\label{fig:uncurated_samples}
\end{figure}
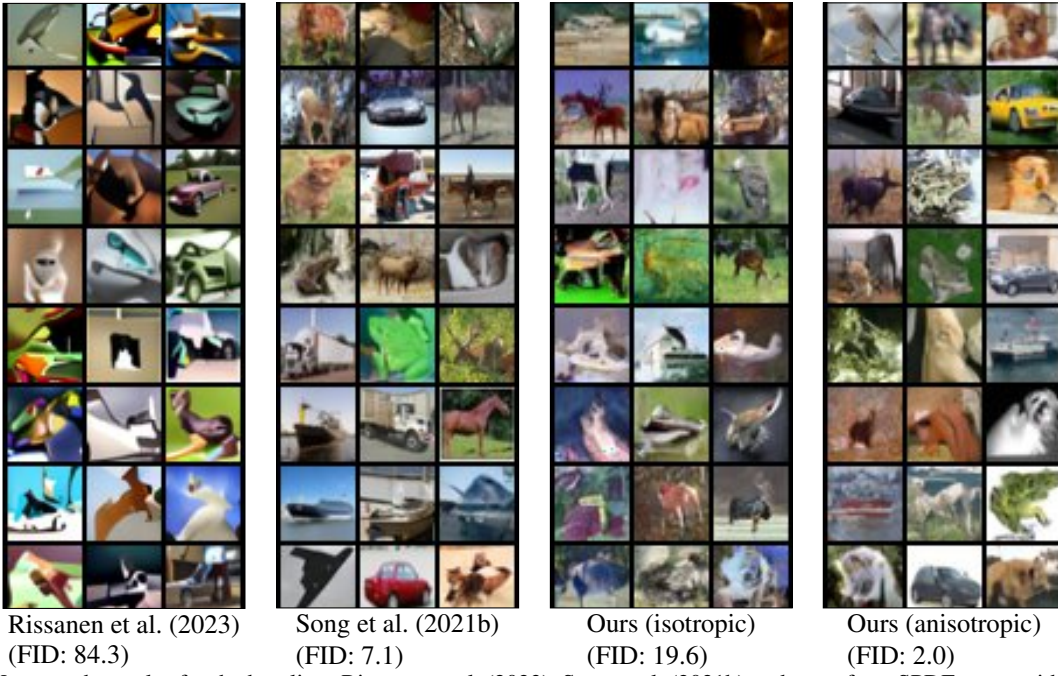

\newpage
\section{Uncurated generated samples on \textsc{\textsc{CelebA}}}\label{sec:uncurated-samples-celeba}

\begin{figure}[htbp]
    \centering
    \includegraphics[width = .75\linewidth]{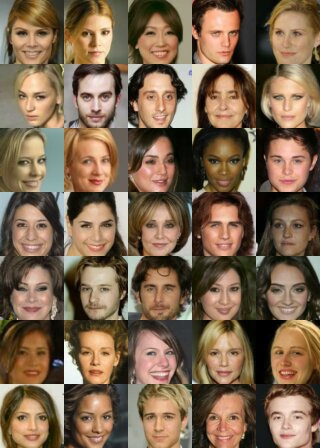}
    \caption{
    Uncurated samples for
    \emph{Ours (anisotropic)} (see \cref{sec:exp_process_aniso_heat_iso_noise}). The generated images are produced by a model trained from scratch \textemdash\ without initialization from a pre-trained network \textemdash\ for 200{,}000 iterations.}
\end{figure}
\begin{figure}[htbp]
    \centering
    \includegraphics[width = .75\linewidth]{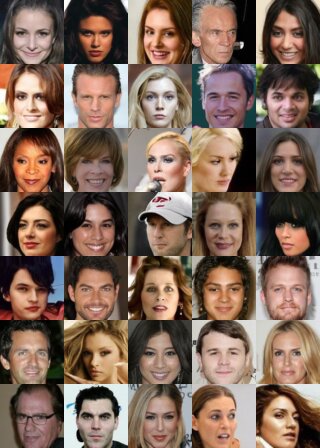}
    \caption{
    Uncurated samples for \citet{song2021scorebased}. The generated images are produced from the checkpoint provided by the authors.}
\end{figure}
\begin{figure}[htbp]
    \centering
    \includegraphics[width = .75\linewidth]{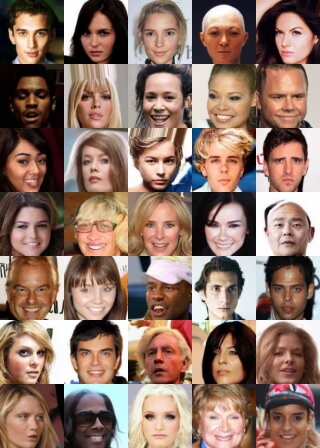}
    \caption{
    Uncurated samples for \citet{lipman2022flow}. The generated images are produced by a model trained from scratch \textemdash\ without initialization from a pre-trained network \textemdash\ for 200{,}000 iterations.}
\end{figure}

\newpage
\section{Uncurated generated samples on \textsc{ImageNet2012}}\label{sec:uncurated-samples-imagenet2012}

\begin{figure}[htbp]
    \centering
    \includegraphics[width = .75\linewidth]{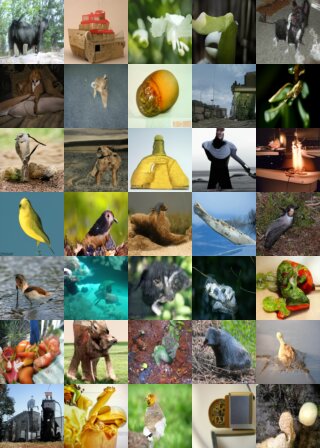}
    \caption{
    Uncurated samples for
    \emph{Ours (anisotropic)} (see \cref{sec:exp_process_aniso_heat_iso_noise}). The generated images are produced by a model trained from scratch \textemdash\ without initialization from a pre-trained network \textemdash\ for 200{,}000 iterations.}
\end{figure}
\begin{figure}[htbp]
    \centering
    \includegraphics[width = .75\linewidth]{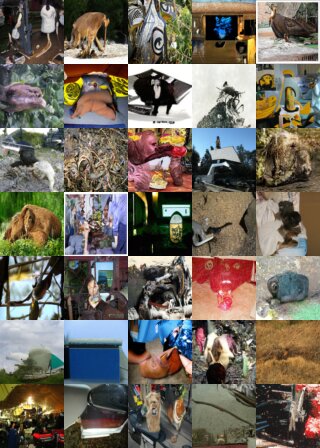}
    \caption{
    Uncurated samples for \citet{song2021scorebased}. The generated images are produced from the checkpoint provided by the authors.}
\end{figure}
\begin{figure}[htbp]
    \centering
    \includegraphics[width = .75\linewidth]{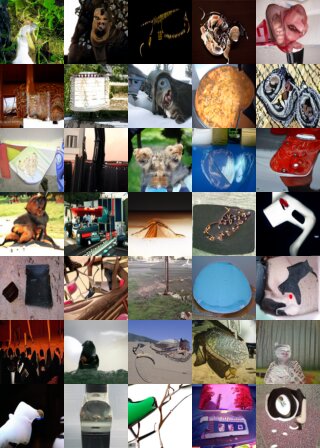}
    \caption{
    Uncurated samples for \citet{lipman2022flow}. The generated images are produced by a model trained from scratch \textemdash\ without initialization from a pre-trained network \textemdash\ for 200{,}000 iterations.}
\end{figure}

\newpage
\section{Uncurated generated samples on \textsc{LSUN\slash church\_outdoor}}\label{sec:uncurated-samples-lsun}

\begin{figure}[htbp]
    \centering
    \includegraphics[width = .65\linewidth]{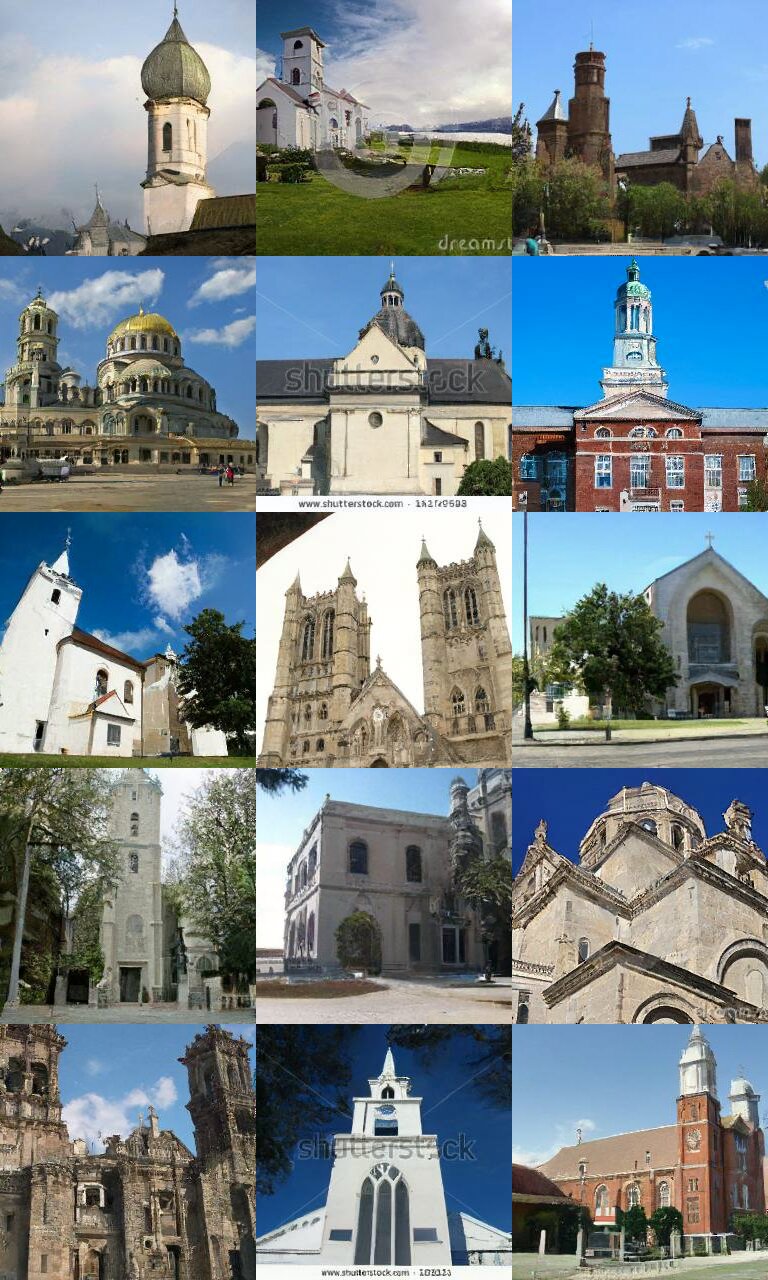}
    \caption{
    Uncurated samples for \emph{Ours (anisotropic)} (see \cref{sec:exp_process_aniso_heat_iso_noise}). The generated images are produced by a model trained from scratch \textemdash\ without initialization from a pre-trained network \textemdash\ for 200{,}000 iterations.}
\end{figure}
\begin{figure}[htbp]
    \centering
    \includegraphics[width = .65\linewidth]{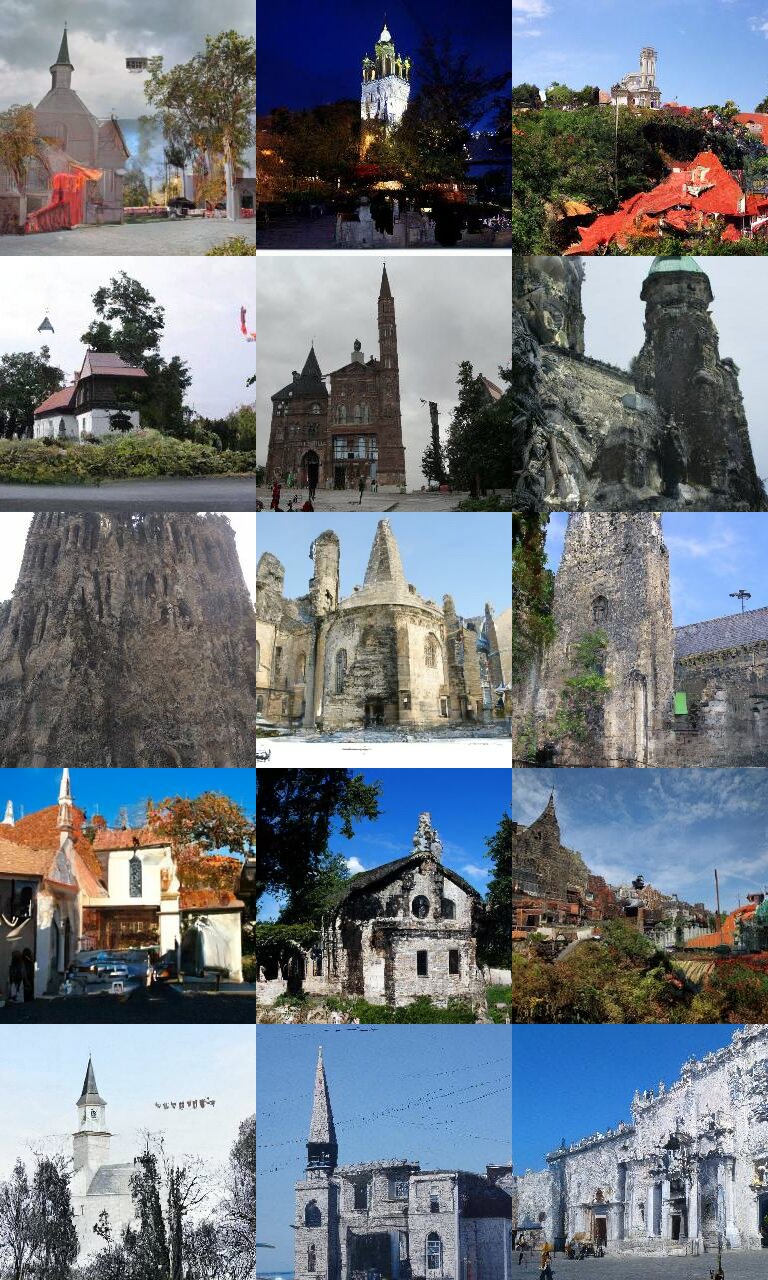}
    \caption{
    Uncurated samples for \citet{song2021scorebased}. The generated images are produced from the checkpoint provided by the authors.}
\end{figure}

\newpage
\section{Uncurated generated samples on \textsc{LSUN\slash bedroom}}\label{sec:uncurated-samples-bedroom}

\begin{figure}[htbp]
    \centering
    \includegraphics[width = .65\linewidth]{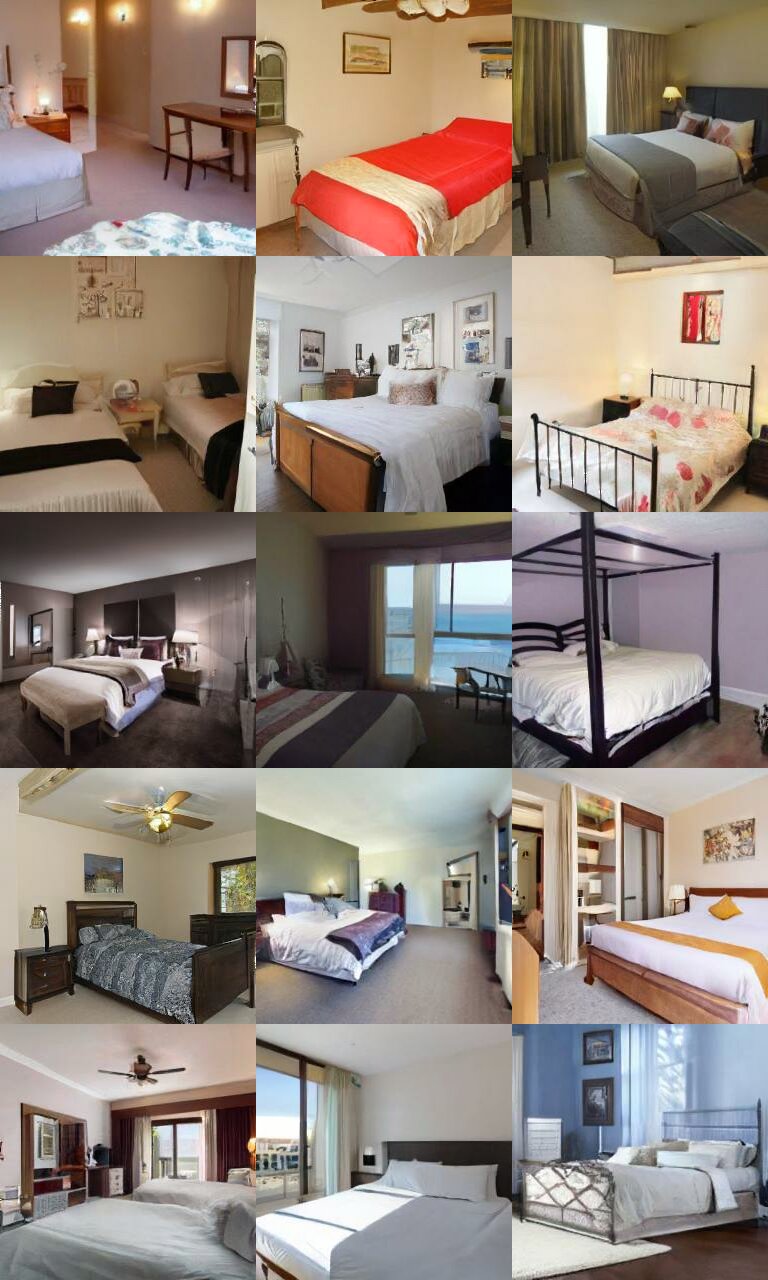}
    \caption{
    Uncurated samples for \emph{Ours (anisotropic)} (see \cref{sec:exp_process_aniso_heat_iso_noise}). The generated images are produced by a model trained from scratch \textemdash\ without initialization from a pre-trained network \textemdash\ for 200{,}000 iterations.}
\end{figure}
\begin{figure}[htbp]
    \centering
    \includegraphics[width = .65\linewidth]{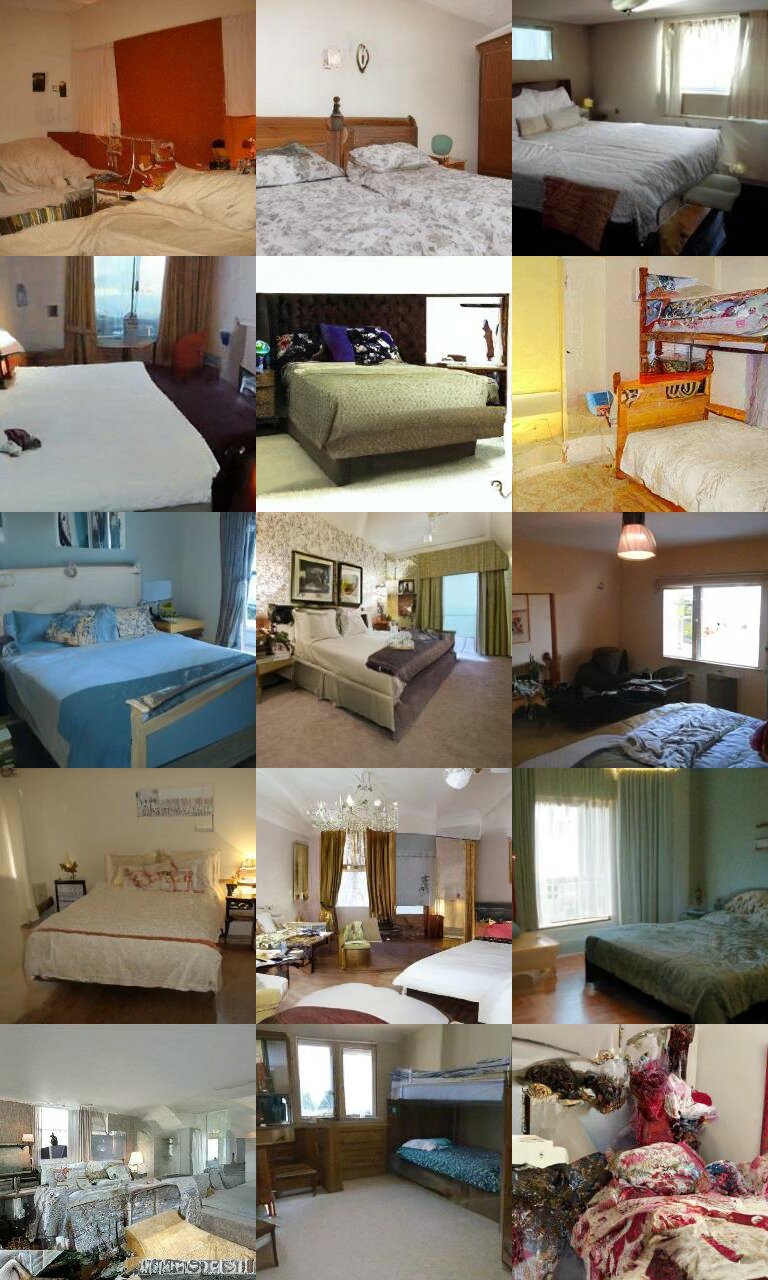}
    \caption{
    Uncurated samples for \citet{song2021scorebased}. The generated images are produced from the checkpoint provided by the authors.}
\end{figure}

\newpage
\section{Uncurated generated samples on \textsc{LSUN\slash church\_outdoor} from our latent space experiment}\label{sec:uncurated-samples-lsun-latent}

\begin{figure}[htbp]
    \centering
    \includegraphics[width = .65\linewidth]{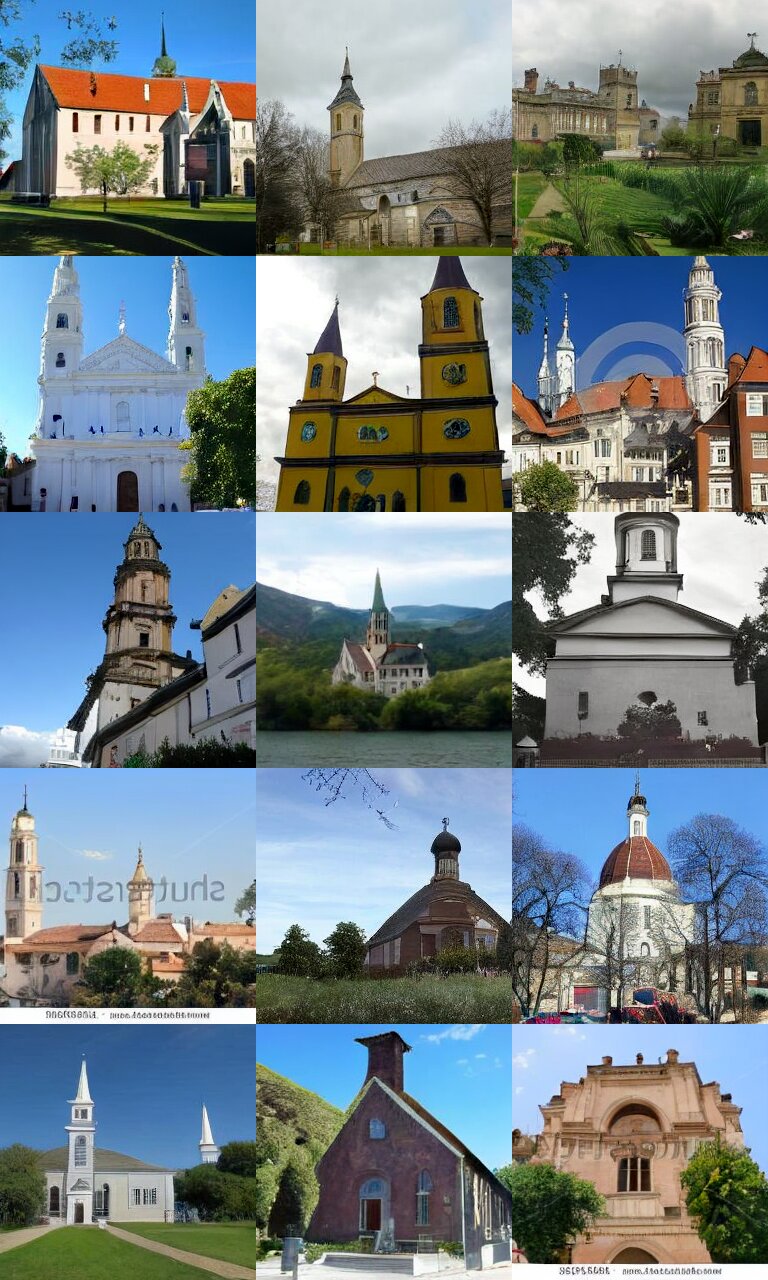}
    \caption{
    Uncurated samples for \emph{Ours (anisotropic)} (see \cref{sec:exp_process_aniso_heat_iso_noise}) from our latent space experiment. The generated images are produced by a model trained from scratch for 200{,}000 iterations. Watermarks are part of the original dataset.}
\end{figure}

\newpage
\section{Uncurated generated samples on \textsc{LSUN\slash bedroom} from our latent space experiment}\label{sec:uncurated-samples-bedroom-latent}

\begin{figure}[htbp]
    \centering
    \includegraphics[width = .65\linewidth]{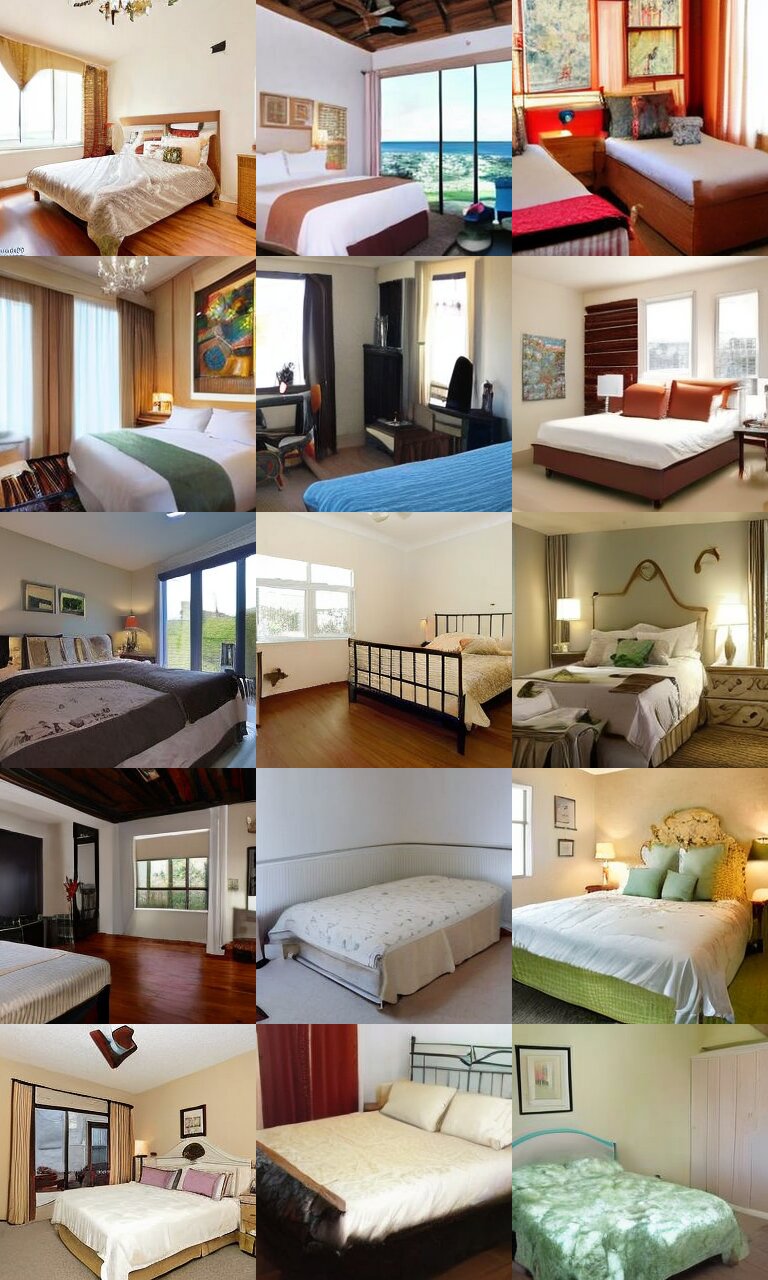}
    \caption{
    Uncurated samples for \emph{Ours (anisotropic)} (see \cref{sec:exp_process_aniso_heat_iso_noise}). The generated images are produced by a model trained from scratch --- without initialization from a pre-trained network --- for 200{,}000 iterations.}
\end{figure}

\end{document}